\documentclass[altaffilletter, aps, pra, showpacs, twocolumn, amsfonts, amsmath, amssymb, floatfix, groupedaddress]{revtex4}

\usepackage{graphicx}
\usepackage{dcolumn}
\usepackage{bm}
\usepackage{hyperref}

\begin{document}

\title{Fast compression of a cold atomic cloud using a blue detuned crossed dipole trap}

\author{Tom Bienaim\'e}
\altaffiliation[Present address: ]{Niels Bohr Institute, University of Copenhagen, Blegdamsvej 17, DK-2100 Copenhagen, Denmark}
\author{Giovanni Barontini}
\altaffiliation[Present address: ]{Research Center OPTIMAS, Technische Universit\"at Kaiserslautern, DE-67663 Kaiserslautern, Germany}
\author{Laure Mercier de L\'epinay}
\author{Louis Bellando}
\author{Julien Chab\'e}
\author{Robin Kaiser}

\affiliation{Universit\'e de Nice Sophia Antipolis, CNRS, Institut Non-Lin\'eaire de Nice, UMR 7335, F-06560 Valbonne, France}

\date{\today}

\begin{abstract}
We present the experimental realization of a compressible blue detuned crossed dipole trap for cold atoms allowing for fast dynamical compression ($\sim 5 - 10 \, \text{ms}$) of $5 \times 10^7$ Rubidium atoms up to densities of $\sim 10^{13} \, \text{cm}^{-3}$. The dipole trap consists of two intersecting tubes of blue-detuned laser light. These tubes are formed using a single, rapidly rotating laser beam which, for sufficiently fast rotation frequencies, can be accurately described by a quasi-static potential. The atomic cloud is compressed by dynamically reducing the trap volume leading to densities close to the Ioffe-Reggel criterion for light localization.

\end{abstract}


\maketitle


\section{Introduction}

The use of optical dipole traps for manipulating and trapping ultra-cold atoms has been crucial to the evolution of this field of research. For example they can be used to form a Bose-Einstein condensate (BEC) \cite{Chapman01}, create artificial crystal of light \cite{Bloch02} or study physics in low dimensions by freezing spatial degrees of freedom \cite{Dalibard06}. 

Light-matter interactions in dense regime is a very dynamic and challenging field of research \cite{Akkermans08,Kaiser09,Sokolov09} where one of the long standing problem is the understanding of the role of cooperative effects (superradiance \cite{Scully06, Bienaime10}, subradiance \cite{Bienaime12}, collective Lamb shift \cite{Scully09}) and disorder (weak \cite{Labeyrie99} and strong localization \cite{Anderson58}). In order to reach this regime, a dipole trap can be used to compress the cloud to high densities where the strong localization phase transition \cite{Anderson58} is expected to occur at a threshold given by the Ioffe-Reggel criterion \cite{Ioffe60} $k \cdot l \sim 1$, where $k = 2 \pi / \lambda$ is the light wavevector and $l=1/(n \sigma)$ is the mean free path ($n$ the atomic density, $\sigma$ the scattering cross section). For resonant two-level systems, the scattering cross section $\sigma_0 = 3 \lambda^2 / (2 \pi)$ allows to express the Ioffe-Reggel criterion as $n \lambda^3 \sim 1$. Such densities correspond for Rubidium atoms to $10^{13} - 10^{14} \, \text{cm}^{-3}$, three orders of magnitude higher than magneto-optical trap (MOT) densities. These high densities are commonly obtained in dipole trap for bosonic \cite{Chapman01} or fermionic \cite{Thomas02} ultra-cold gases. 
However, the relatively low number of atoms ($\sim 10^5$) and long duty cycles ($\sim 10 \, \text{s}$) make it difficult to efficiently study light-matter interaction in dense regimes \cite{Balik09} where a large number of atoms as well as short duty cycles are important assets for efficient detection of signatures of cooperative effects and/or strong localization of light. Indeed the Ioffe-Reggel criterion should not be confused with the BEC threshold $n \Lambda_T^3 \sim 1$, where $\Lambda_T$ is the thermal de Broglie wavelength. Contrary to Bose-Einstein condensation, we do not expect a drastic constraint on temperature for the Ioffe-Regel criterion of strong localization.

In this paper, we present a compressible blue detuned crossed dipole trap to achieve a very fast dynamical compression ($\sim 5 - 10 \, \text{ms}$) of a large number of $^{87}$Rb atoms ($\sim 5 \times 10^7$) to densities compatible with the Ioffe-Reggel criterion, i.e. $\sim 10^{13} \, \text{cm}^{-3}$. Trapping atoms in a `dark' region surrounded by blue-detuned light has several advantages such as minimizing photon scattering, light shifts of the atomic levels, and light-assisted collisional losses \cite{Grimm99}. Experimentally, blue-detuned trap are more difficult to produce than red ones. However, an original method \cite{Davidson00} consists in using a focused Gaussian laser beam which is rapidly rotating using two perpendicular acousto-optical modulators (AOMs). If the rotation frequency is sufficiently high, the resulting time-averaged potential forms a tube of light. Crossing two of these tubes leads to a `dark' volume where atoms are trapped. This method allows to dynamically control the shape and the size of the trap which for example might be used to optimize the loading efficiency with a large trapping volume and then compress the cloud using a fast dynamical reduction of the trap size. These blue detuned time-averaged potentials have been used in the past to study thermal cloud compression \cite{Davidson00}, optical billiards and chaos \cite{Milner01,Friedman01,Kaplan01}, or designing microscopically tailored potentials for BECs \cite{Henderson09} or ultra-cold Fermi gases \cite{Zimmermann11}. For red dipole traps, a dynamical compression allowed reaching quantum degeneracy \emph{via} a runaway evaporative cooling using a mobile lens to change the trap waist dynamically \cite{{Weiss05}}.

\section{Experimental setup}    \label{expsetup}

\subsection{Trap configuration} \label{TrapConf}

\begin{figure*}[t!]
\centering{\includegraphics[height=5cm]{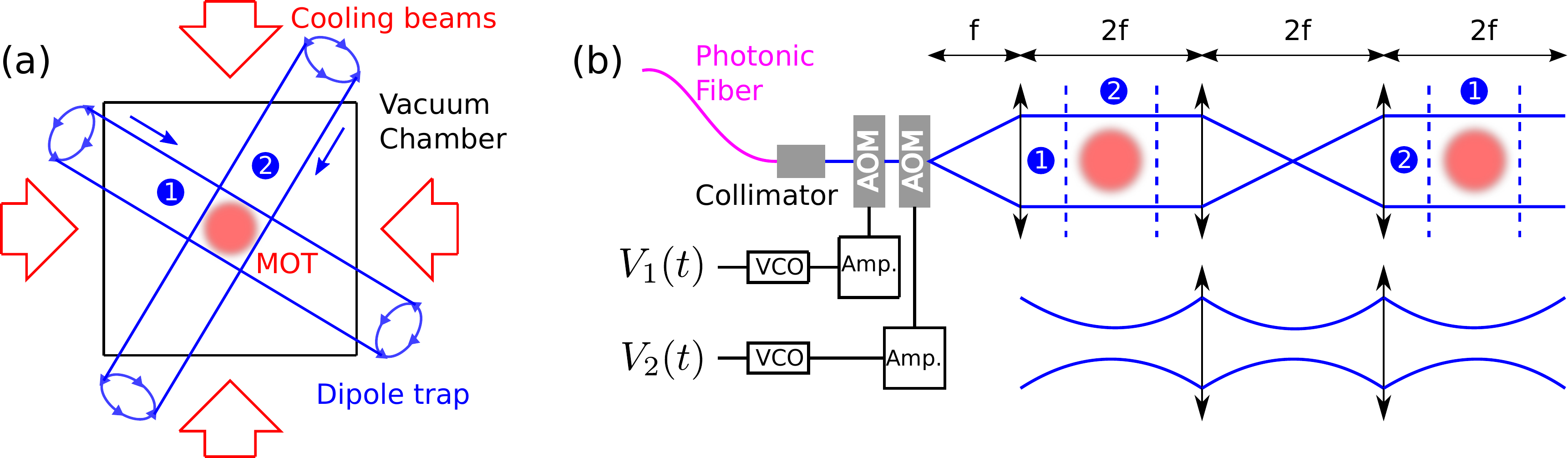}}
\caption{(Color online) Sketch of the experimental setup. (a) Top view of the setup, showing the location of the two horizontal tubes forming the crossed blue detuned dipole trap. (b) Details of the setup used to produce the two tubes ($(+1,+1)$ orders of the AOMs). On the top, only the contour of the trap is sketched while on the bottom only the beam profile is shown. For simplicity, we do not show the two mirrors that are used to cross the tubes 1 and 2 (the dashed lines show the position of the crossing).}
\label{fig1}
\end{figure*}

The trap consists in two tubes of blue-detuned light, crossed at $90^\circ$ to create a box of light (see Fig. \ref{fig1} (a)) where the atoms are confined. The size and the shape of the box can be dynamically adjusted. Fig. \ref{fig1} (b) shows the experimental setup to create the tubes of light. A laser beam with $1 \, \text{mm}$ waist passes through two AOMs (Gooch \& Housego M080-2B/F-GH2), crossed at $90^\circ$. The AOMs are powered by two radio-frequency (RF) signals whose instantaneous frequencies are respectively given by $f_1(t) = f_0 + \Delta f \cos (2 \pi f_m t)$ and $f_2(t) = f_0 + \Delta f \sin (2 \pi f_m t)$. The central frequency $f_0$ is fixed at $80 \, \text{MHz}$, the modulation frequency amplitude $\Delta f$ is at most $20 \, \text{MHz}$ and the modulation frequency $f_m$ is generally set to $90 \, \text{kHz}$. In the $(+1,+1)$ diffraction order of the AOMs, the laser beam is rotating at frequency $f_m$ and a lens, placed at a focal distance of $150 \, \text{mm}$ from the AOMs, creates a time-averaged tube of light with a diameter of $1 \, \text{mm}$ and a waist of $65 \, \mu\text{m}$ in the lens focal plane where the atoms are trapped. Using a system of lenses and mirrors, the tube can be `recycled' and crossed at $90^\circ$ from its initial direction (see Fig. \ref{fig1}). Varying the modulation frequency amplitude $\Delta f$, the trap size can be dynamically controlled.

\begin{figure}[t]
\centering{\includegraphics[height=4.2cm]{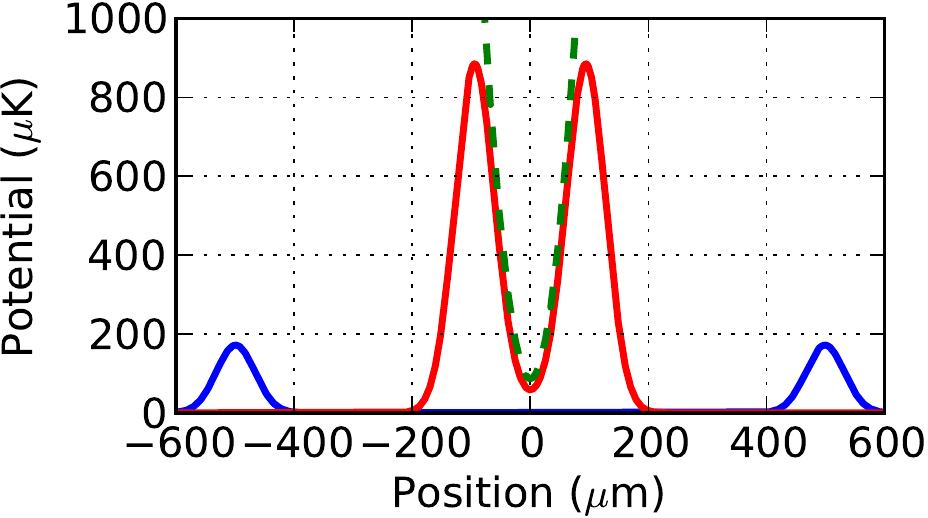}}
\caption{(Color online) Radial profiles of the time-averaged potential for two trap radii. The solid curves correspond to Eq. (\ref{radialformula}) and Eq. (\ref{dip_pot}) for trap radii $a=500 \, \mu\text{m}$ (blue) and $a=100 \, \mu\text{m}$ (red). The green dashed line represents the harmonic approximation of the potential (see Eq. (\ref{POTHARM})) for $a=100 \, \mu\text{m}$. It gives $U_1=85 \, \mu\text{K}$ and $\omega/(2\pi) = 860 \, \text{Hz}$.  Parameters: $P = 200 \, \text{mW}$, $\Delta = 40 \, \text{GHz}$, $w = 65 \, \mu \text{m}$.}
\label{fig2}
\end{figure}

A rotating laser beam with power $P$, waist $w$ and a radial intensity profile given by $I(x,y) = \frac{ 2 P }{\pi w^2} \exp \left(  -2 (x^2 + y^2) / w^2  \right)$ describing a circle of radius $a$, creates a time average intensity profile
\begin{equation}
I(r) = \frac{2P}{\pi w^2} \exp \left( \frac{-2 (r^2 + a^2)}{w^2} \right) \mathcal{I}_0 \left(  \frac{4 a r}{w^2}  \right), \label{radialformula}
\end{equation}
where $r$ is the radial distance of the polar coordinate system and $\mathcal{I}_0$ is the zeroth-order modified Bessel function. For $^{87}$Rb atoms, the dipole potential for a linearly polarized light with detuning $\Delta$ with respect to the D$_2$ line such that $\Delta'_{\text{FS}}   \gg |\Delta| \gg \Delta'_{\text{HFS}}$, where $\hbar \Delta'_{\text{FS}}$ and $\hbar \Delta'_{\text{HFS}}$ are respectively the energy splitting of the fine and hyperfine excited states, is given by \cite{Grimm99}
\begin{equation}
U(r) = \frac{2}{3} \times \frac{\hbar \Gamma^2}{8 \Delta} \frac{I(r)}{I_{\text{sat}}}, \label{dip_pot}
\end{equation}
$\Gamma$ being the linewidth and $I_{\text{sat}}$ the saturation intensity of the transition. For large trap size $a/w \gg 1$, i.e. when the radius is large compared to the waist, Eq. (\ref{dip_pot}) simplifies to the intuitive formula
\begin{equation}
U(r) = U_0 \exp \left( - \frac{2 (r - a)^2 }{ w^2} \right), \label{dip_pot_big}
\end{equation}
where the potential height is given by
\begin{equation}
U_0 = \frac{2}{3} \times \frac{\hbar \Gamma}{8 \sqrt 2 \pi^{3/2} }  \frac{\Gamma}{\Delta} \frac{P}{I_{\text{sat}} a w}. \label{maxpotbar}
\end{equation}
When the trap is small i.e. $a/w < 1$, the two walls of the tube start touching each other and the trap can no longer be described as a box with Gaussian walls but is well approximated by a harmonic potential 
\begin{equation}
U(r) = U_1 + \frac{1}{2} m \omega^2 r^2, \label{POTHARM}
\end{equation} 
where $U_1 = 2/3 \times \frac{\hbar \Gamma}{4 \pi } \frac{\Gamma}{\Delta} \frac{P}{I_{\text{sat}}w^2} \exp \left( \frac{-2a^2}{w^2} \right)$ is the `offset' value of the potential at the center of the trap and $\omega = 2 \sqrt{ \frac{U_1}{m}} \frac{\sqrt{2 a^2 - w^2}}{w^2}$ is the trap frequency. Fig. \ref{fig2} shows the radial profile of the timed-averaged potential for two trap radii as well as its harmonic approximation. The possibility to control the trap frequency by tuning its radius can allow for runaway evaporative cooling by compensating the reduction of the trap frequency due to the lowering of the potential barrier (reducing the intensity or increasing the detuning) by a reduction of the size of the trap. The trap frequency is very sensitive to the trap radius. A careful control of the approach of the final trap radius is thus required when reaching the harmonic regime.

\subsection{Laser system}

The laser system consists of a distributed feedback (DFB) laser diode injecting a semiconductor laser amplifier SACHER delivering up to $1 \, \text{W}$. After the optical isolator, the beam (power $900 \, \text{mW}$) is coupled in a large core ($10 \, \mu \text{m}$), monomode, polarization maintaining photonic crystal fiber (NKT Photonics LMA-PM-10). The coupling efficiency into the fiber is $60 \%$, limited by the quality of the laser mode of the semiconductor amplifier. 

At the output of the fiber, $550 \, \text{mW}$ of collimated linearly polarized light injects the crossed AOMs, yielding $250 \, \text{mW}$ in the $(+1,+1)$ diffraction order for creating the light tube. The power of the beam can be controlled by adjusting the RF power driving the AOMs. The light frequency $\nu$ can be tuned over $120 \, \text{GHz}$ without mode jump by adjusting the current $I$ of the DFB diode ($\mathrm d \nu / \mathrm d I=2 \, \text{GHz} \, \text{mA}^{-1}$). Typical detunings used with this setup range between $5 \, \text{GHz}$ and $80 \, \text{GHz}$. 

Such semiconductor laser systems have the advantage of moderate costs and the simplicity of implementation. However, they often have a modest mode quality and more importantly they possess an amplified spontaneous emission background which spreads over $40 \, \text{nm}$, containing photons at resonance with the atomic line. For experiments where this background spectrum is a limitation, a titanium-sapphire laser or a frequency doubled laser might be a more convenient choice.

\subsection{Electronic control of RF-signals}

\begin{figure}[t]
\centering{\includegraphics[height=4.2cm]{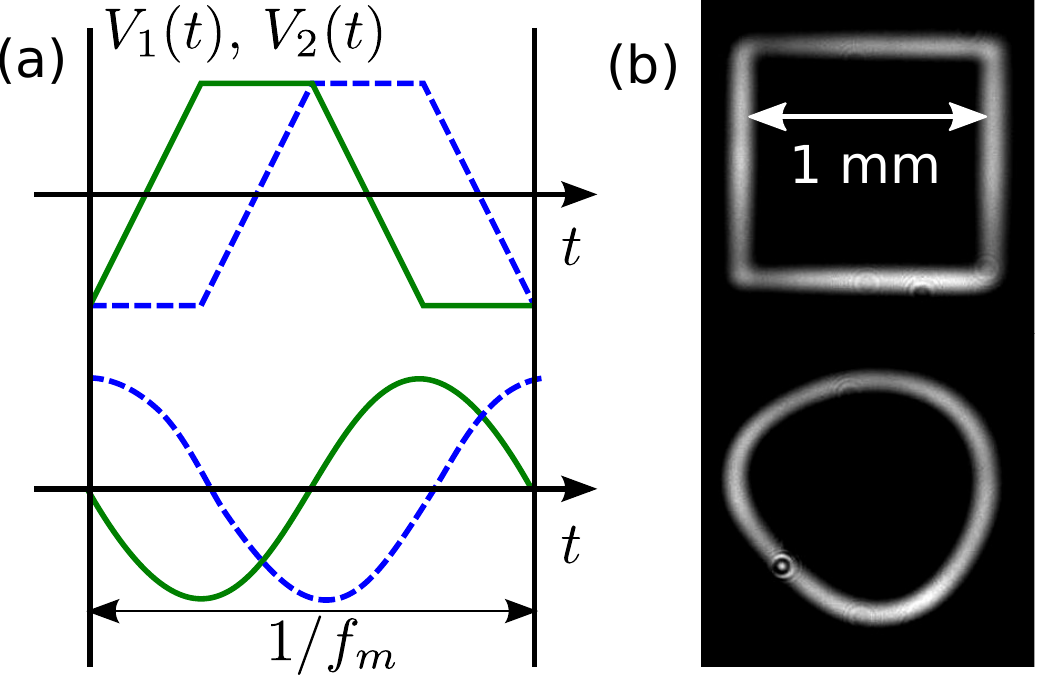}}
\caption{(Color online) Measured cross sections of the blue detuned tubes for two different VCO modulations. As an example, we show two useful configurations: the square and the circle. (a) Signals $(V_1(t), V_2(t))$ that are sent to the VCOs. (b) Pictures of the resulting cross sections taken at the position where the atoms are trapped. The size of the trap is $L \simeq 1 \, \text{mm}$ and the waist $w = 65 \, \mu\text{m}$.}
\label{fig3}
\end{figure}

In order to create the tubes of light, a precise control on the RF-signals feeding the AOMs is necessary. For this, we use voltage control oscillators (VCOs) delivering RF-signals whose instant frequencies $f_i(t)$ linearly depend on the input voltages $V_i(t)$. Thus, the two input voltages $(V_1(t),V_2(t))$ are associated to a position $(x(t),y(t))$ in the $(+1,+1)$ diffraction order of the AOMs, such that it is possible to create tubes of light with arbitrary cross sections. Fig. \ref{fig3} shows the signals $V_i(t)$ which produce circular and square tubes as well as the subsequent experimental pictures of these cross sections. We use homemade VCOs, based on the \emph{Mini-Circuits} POS-150+ chip with output frequency between $50 \, \text{MHz}$ and $150 \, \text{MHz}$ and a $3 \, \text{dB}$ input modulation bandwidth of $100 \, \text{kHz}$. Two phased locked \emph{Agilent} 33220A function generators are used to drive the VCOs. The VCOs are feeding two \emph{Mini-Circuits} ZHL-1A RF amplifiers.

In Fig. \ref{fig3} (b), we notice that the ring-shaped potential is slightly asymmetric. We attribute this to nonlinearities in the overall system response $(x(V_1(t)),y(V_2(t)))$ which is supported by the fact that this asymmetry is reduced for smaller traps since they require lower modulation amplitudes. The square-shaped trap is not affected because its parametric equations do not lead to a combined motion in the $x$ and $y$ directions (the nonlinearities only imply that straight lines are not drawn at constant speed). By designing and engineering the synthesizer signals driving the VCOs, it would be possible to compensate for the trap asymmetry. However, all the measurements performed in this paper use the ring-shaped potential without compensating for the asymmetry to keep the system complexity to a minimum.

\subsection{Parameters and experimental sequence} \label{paramexpseq}

Unless otherwise stated, the parameters of the trap are: size $L = 1 \, \text{mm}$ (radius $a = 500 \, \mu\text{m}$), waist $w = 65 \, \mu\text{m}$, detuning $\Delta = 40 \, \text{GHz}$, rotation frequency $f_m = 90 \, \text{kHz}$, laser power for a single tube $P = 200 \, \text{mW}$, linear polarization. For these parameters, the potential height for $^{87}$Rb atoms is $190 \, \mu\text{K}$ (see Eq. (\ref{maxpotbar})).

\begin{figure}[t]
\centering{\includegraphics[height=4.2cm]{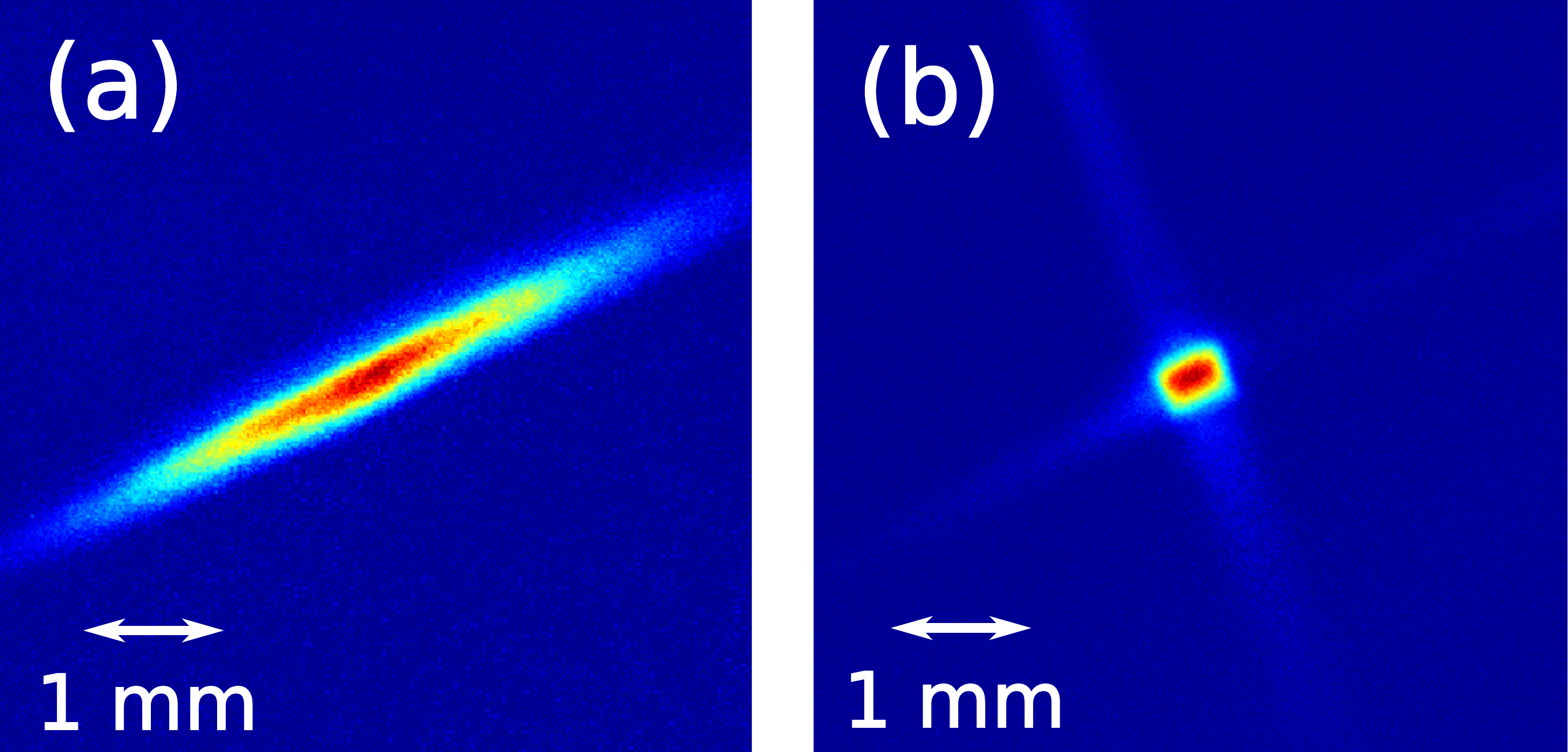}}
\caption{(Color online) \emph{In situ} fluorescence images of the atoms in the dipole trap taken from the top of the cell after $100 \, \text{ms}$ of holding time. (a) Single tube trap. (b) Crossed trap. Parameters: $P = 200 \, \text{mW}$, $\Delta = 40 \, \text{GHz}$, $L = 1 \, \text{mm}$, $w = 65 \, \mu \text{m}$, $f_m = 90 \, \text{kHz}$.}
\label{fig4}
\end{figure}

We first load a MOT of $^{87}$Rb atoms from a vapor cell with a background gas pressure of $\sim 10^{-9} \, \text{mbar}$. All lasers are tuned close to the $\text{D}_2$ line of $^{87}$Rb and are derived from DFB diodes, conveniently amplified with a tapered amplifier and controlled via AOMs. In this series of experiments, we deliberately choose to work with a moderate number of atoms to investigate the performances of the novel trapping scheme. We trap $\sim 5 \times 10^7$ atoms in $2.5 \, \text{s}$. The loading time can be reduced by increasing the Rubidium background pressure when a larger atom number needs to be trapped using e.g. ultraviolet LEDs to temporarly increase the hot gas pressure during the MOT loading - the so-called light-induced atomic desorption (LIAD) \cite{Gozzini93,Anderson01}. The cooling laser detuning is $-3 \, \Gamma$ and the temperature of the cloud is $\sim 55 \, \mu\text{K}$. We then apply a $50 \, \text{ms}$ temporal dark MOT period where the intensity of the repumping laser is reduced by a factor of $10$ and the detuning of the cooling beam is increased from $-3 \, \Gamma$ to $-6 \, \Gamma$. This allows to compress and produce a homogeneous distribution of atoms, mainly in the $F=1$ hyperfine ground state, with a density $\sim 10^{11} \, \text{cm}^{-3}$ and a temperature $\sim 20 \, \mu\text{K}$. During the temporal dark MOT period, the intensity of the dipole trap is progressively ramped up in order to maximize the `mode matching' between the dark MOT and the dipole trap. After this, the magnetic field and the laser cooling beams are turned off. In order to keep the atoms in a single hyperfine level, we choose to keep the repumping laser on at all times after loading, thus forcing the atoms into the $F=2$ hyperfine level \footnote{Pumping the atoms into the $F=1$ hyperfine level would instead require the development of a specific depumping laser. Indeed maintaining the MOT cooling beams on to pump the atoms in the $F=1$ hyperfine level would not be efficient and would lead to several scattered photons, thus preventing us from attaining our goal of reaching high densities.}. Then the atoms evolve freely in the dipole potential and the trapping time varies between $5 \, \text{ms}$ and $1.2 \, \text{s}$. Using absorption or fluorescence imaging techniques, the properties of the cloud e.g. number of atoms, temperature, density are measured. In the following, unless otherwise stated, absorption imaging from the side of the cell is used to perform quantitative measurements. Fig. \ref{fig4} shows \emph{in situ} fluorescence images of the cloud taken from the top of the cell for a single tube and for the crossed dipole trap.

\section{Loading}

The loading period corresponds to the transfer of the atoms from the dark MOT to the dipole trap. During the first $\sim 60 \, \text{ms}$, the atoms are in a transient regime before reaching a quasi-steady regime. In order to optimize the transfer efficiency from the dark MOT to the dipole trap, it is important to understand the mechanisms occurring during this stage.

\subsection{Transient regime}

\begin{figure}[t!]
\centering{\includegraphics[height=4.2cm]{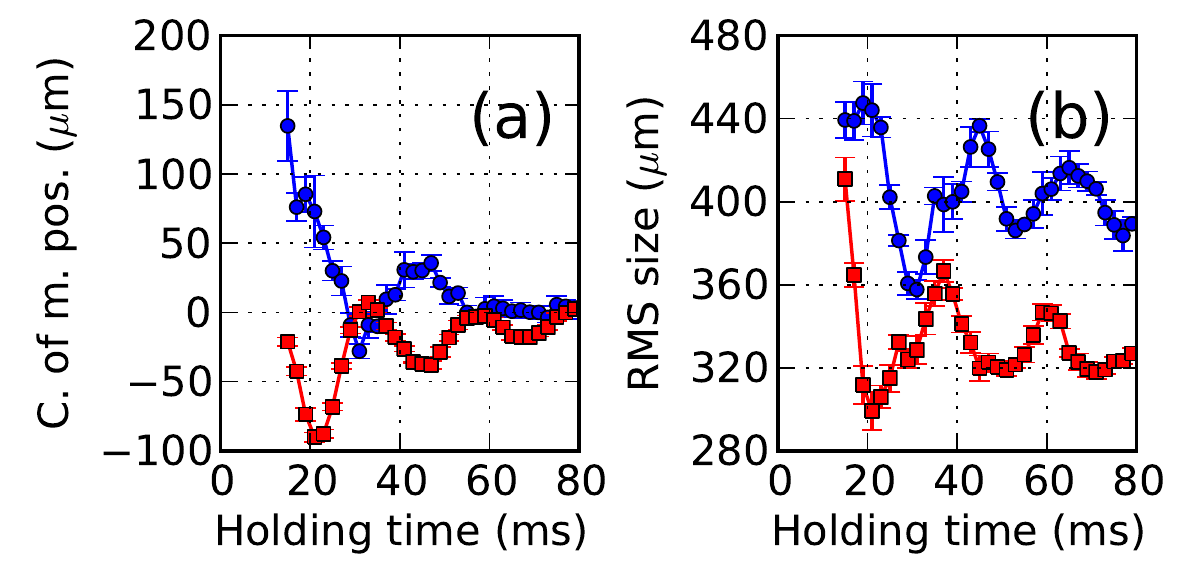}}
\caption{(Color online) Center of mass position (a) and RMS size (b) of the cloud after a $5 \, \text{ms}$ time of flight as a function of the holding time in the dipole trap. The blue points correspond to measurements along the $x$-axis which is orthogonal to the direction of gravity. The red squares represent measurements along the $z$-axis, which is defined to point opposite to gravity. This figure shows center of mass oscillations and breathing modes during the transient regime. Parameters: $P = 200 \, \text{mW}$, $\Delta = 40 \, \text{GHz}$, $L = 1 \, \text{mm}$, $w = 65 \, \mu \text{m}$, $f_m = 90 \, \text{kHz}$.}
\label{fig5}
\end{figure}

During the transfer of the atoms from the dark MOT to the dipole trap, the `mode matching' is not perfect exciting breathing modes and oscillations of the center of mass of the cloud. Fig. \ref{fig5} shows the center of mass position and the root mean square (RMS) size of the cloud after a $5 \, \text{ms}$ time of flight as a function of the holding time in the dipole trap. We notice that the cloud is squeezed along the direction of gravity (smaller RMS size on Fig. \ref{fig5} (b)) as can also be seen on the \emph{in situ} images (see Fig. \ref{fig7}). We observe oscillations of the cloud size that we identify as \emph{breathing modes}. Their period is similar to the one observed on the center of mass position. The oscillation period is $\sim 25 \, \text{ms}$ which is compatible with atoms falling from a height $h = 500 \, \mu\text{m}$ and bouncing on the bottom of the trap with a period $2 \sqrt{2h/g} \sim 20 \, \text{ms}$. The damping of the oscillations is important because of the strong trap anharmonicity. After $\sim 60 \, \text{ms}$, the oscillations are almost completely damped and the trap enters the so-called \emph{quasi-steady regime}.

\subsection{Trapped atom number}

\begin{figure}[t!]
\centering{\includegraphics[height=4.2cm]{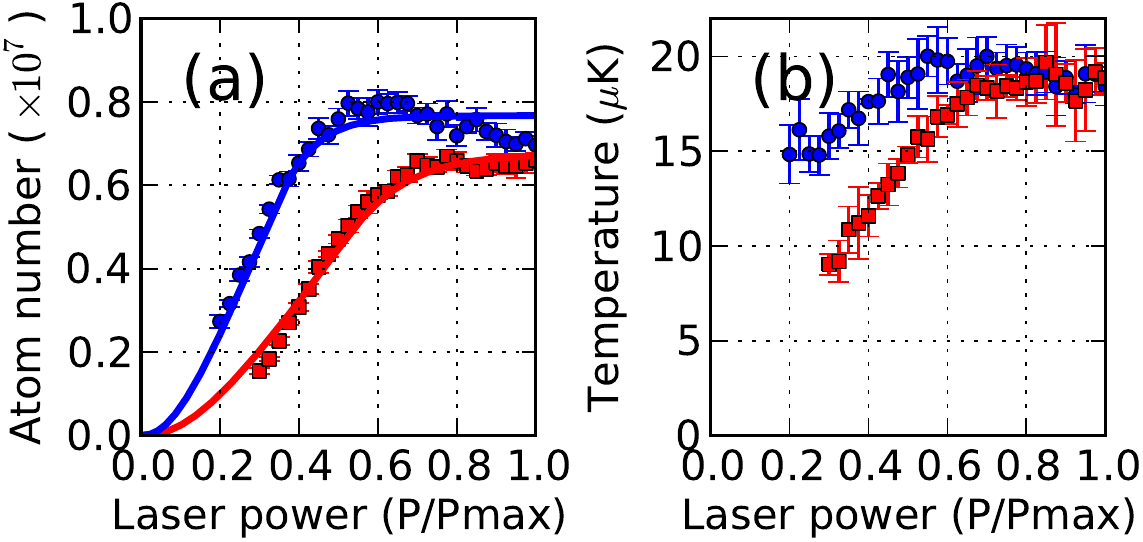}}
\caption{(Color online) Number of atoms (a) and cloud temperature (b) measured $60 \, \text{ms}$ after loading as a function of the laser power ($P_{\text{max}} = 200 \, \text{mW}$). The blue points correspond to a detuning of $20 \, \text{GHz}$ and the red squares to $\Delta = 40 \, \text{GHz}$. The solid lines on (a) are fits of the data according to Eq. (\ref{fraction}) using the potential barrier height $U$ as the only free fitting parameter. We obtain $U = 320 \, \mu\text{K}$ for $\Delta = 20 \, \text{GHz}$ and $U = 190 \, \mu\text{K}$ for $\Delta = 40 \, \text{GHz}$. Parameters: $L = 1 \, \text{mm}$,  $w = 65 \, \mu \text{m}$, $f_m = 90 \, \text{kHz}$.}
\label{fig6}
\end{figure}

At the end of the transient regime, $\sim 60 \, \text{ms}$ after the atom transfer from the dark MOT, we study the number of trapped atoms and the temperature of the cloud as a function of the laser power which can be adjusted between $0$ and $200 \, \text{mW}$. Fig. \ref{fig6} shows the data for two different laser detuning $\Delta = 20 \, \text{GHz}$ (blue points) and $\Delta = 40 \, \text{GHz}$ (red squares). The trap is loaded from a dark MOT with $\simeq 1.3\times 10^7$ atoms and a temperature of $\simeq 22 \, \mu\text{K}$. The size of the trap is $L \simeq 1 \, \text{mm}$. 

Both the number of atoms and temperature reach a plateau after $P = 0.5 \, P_{\text{max}}$ for $\Delta = 20 \, \text{GHz}$ and $P = 0.7 \, P_{\text{max}}$ for $\Delta = 40 \, \text{GHz}$. The temperature plateau corresponds to the temperature of the dark MOT. This can be understood by noticing that when the barrier height is sufficiently high, all the atoms (initially into the trapping region) are trapped and their temperature is the one of the dark MOT. Before the plateau, the temperature increases linearly with laser power which can be understood as a consequence of the linear increase of the potential barrier with laser power. The red squares in Fig. \ref{fig6} (b) clearly point at the origin when $P \rightarrow 0$, which is consistent with the fact that only atoms with an energy (kinetic plus potential) smaller than the barrier height are trapped during loading. However, the blue points do not extend to the origin when $P \rightarrow 0$, which might be due to spontaneous emission heating that is more important for a detuning of $20 \, \text{GHz}$ than for $40 \, \text{GHz}$.

We define the loading efficiency as the number of atoms into the trap after $60 \, \text{ms}$ of holding time (i.e. at the end of the transient regime) divided by the number of atoms in the dark MOT measured before loading. For the data presented on Fig. \ref{fig6}, the loading efficiency is $60 \, \%$ which corresponds to an excellent value compared to what is usually observed in red detuned dipole traps. When the size of the dark MOT is smaller than the trap volume,  the non-trapped atoms correspond to those having a too large energy (kinetic plus potential): they `jump' over the potential barrier. In case gravity is compensated (by e.g. optical pumping of the atoms in a particular Zeeman sublevel and applying a vertical magnetic field gradient), the atoms would only possess kinetic energy which should significantly improve the loading efficiency. The maximum estimated number of atoms that can be trapped is $\sim 10^8$ (equal to the density of the dark MOT $\sim 10^{11} \, \text{cm}^{-3}$, times the trapping volume $\sim 1 \, \text{mm}^3$).

\subsubsection*{Model for trap loading}
We consider the case of a box-shaped trap and suppose that in the dark MOT, the atoms are uniformly distributed in space, with a momentum probability distribution given by the Boltzmann distribution \begin{equation*}
p(\mathbf p,\mathbf r) = \frac{\Lambda_T^3}{L^3} \exp \left[ - \frac{\mathbf p^2}{2 m k_B T} \right],
\end{equation*}
where $\Lambda_T = h/\sqrt{2 \pi m k_B T}$ is the thermal wavelength. The probability density for an atom to have an energy $E$ is
$p(E) = \int \frac{\mathrm d^3 \mathbf p \, \mathrm d^3 \mathbf r}{h^3} \delta(E(\mathbf p, \mathbf r) - E) \, p(\mathbf p,\mathbf r)$, where the energy of the atom takes the form $E(\mathbf p,\mathbf r) =  \mathbf p^2/ (2 m) + m g z$, where the $z$-axis is defined along the direction opposite to gravity. After some calculations, we find
\begin{equation}
p(E) = \frac{2}{\sqrt \pi k_B T} \sqrt{\frac{E}{k_B T}} \exp \left[ - \frac{E}{k_B T} \right]  \mathcal F \left(\frac{E}{mgL} , \frac{mgL}{k_B T} \right), \label{pE}
\end{equation}
where the function $\mathcal F$ is defined as $\mathcal F (\alpha,\beta) = \int_0^{\min (\alpha,1) } \mathrm d u \, \sqrt{1 - u/\alpha} \, \exp (\beta u)$. In the $g \rightarrow 0$ case, i.e. without gravity, Eq. (\ref{pE}) simplifies to the well-known free space density probability $p(E) = 2 / (\sqrt \pi k_B T) \sqrt{E / (k_B T)} \exp \left[ - E / (k_B T) \right]$, where in front of the Boltzmann factor $\exp \left[ - E / (k_B T) \right]$, one recognizes the three dimensional free space density of states.
If the height of the potential barrier is $U$, then the fraction of atoms $N/N_0$ that are trapped during loading is given by  
\begin{equation}
\frac{N}{N_0} = \int_0^U \mathrm d E \, p(E). \label{fraction}
\end{equation}
The solid lines in Fig. \ref{fig6} correspond to fits according to Eq. (\ref{fraction}) which allows to determine, using a single free fitting parameter, the potential height $U$ corresponding to $P_{\text{max}} = 200 \, \text{mW}$. We obtain $U = 320 \, \mu\text{K}$ for $\Delta = 20 \, \text{GHz}$ and $U = 190 \, \mu\text{K}$ for $\Delta = 40 \, \text{GHz}$. We can compare these values to theoretical estimations by substituting the experimental parameters into Eq. (\ref{maxpotbar}) leading respectively to $U = 380 \, \mu\text{K}$ and $U = 190 \, \mu\text{K}$, in good agreement with the values extracted from the fit.

\section{Quasi-steady regime}

After the transient regime leading to the loading of the trap, the evolution of the cloud properties occurs on a longer timescale: the \emph{quasi-steady regime}. In this section, we study the evolution of the cloud in this quasi-steady regime in order to characterize the trap and to understand the mechanisms limiting its performances. To this end, two quantities are of particular interest: the trap lifetime and the temperature of the cloud. The lifetime $\tau$ is determined from an exponential fit $N_0 \exp (-t / \tau)$ to the decay curve of the number of atoms in the trap as a function of the holding time while the temperature is inferred from the RMS size of the cloud measured in time of flight.

\subsection{\emph{In situ} density profile}

\begin{figure}[t]
\centering{\includegraphics[height=4.2cm]{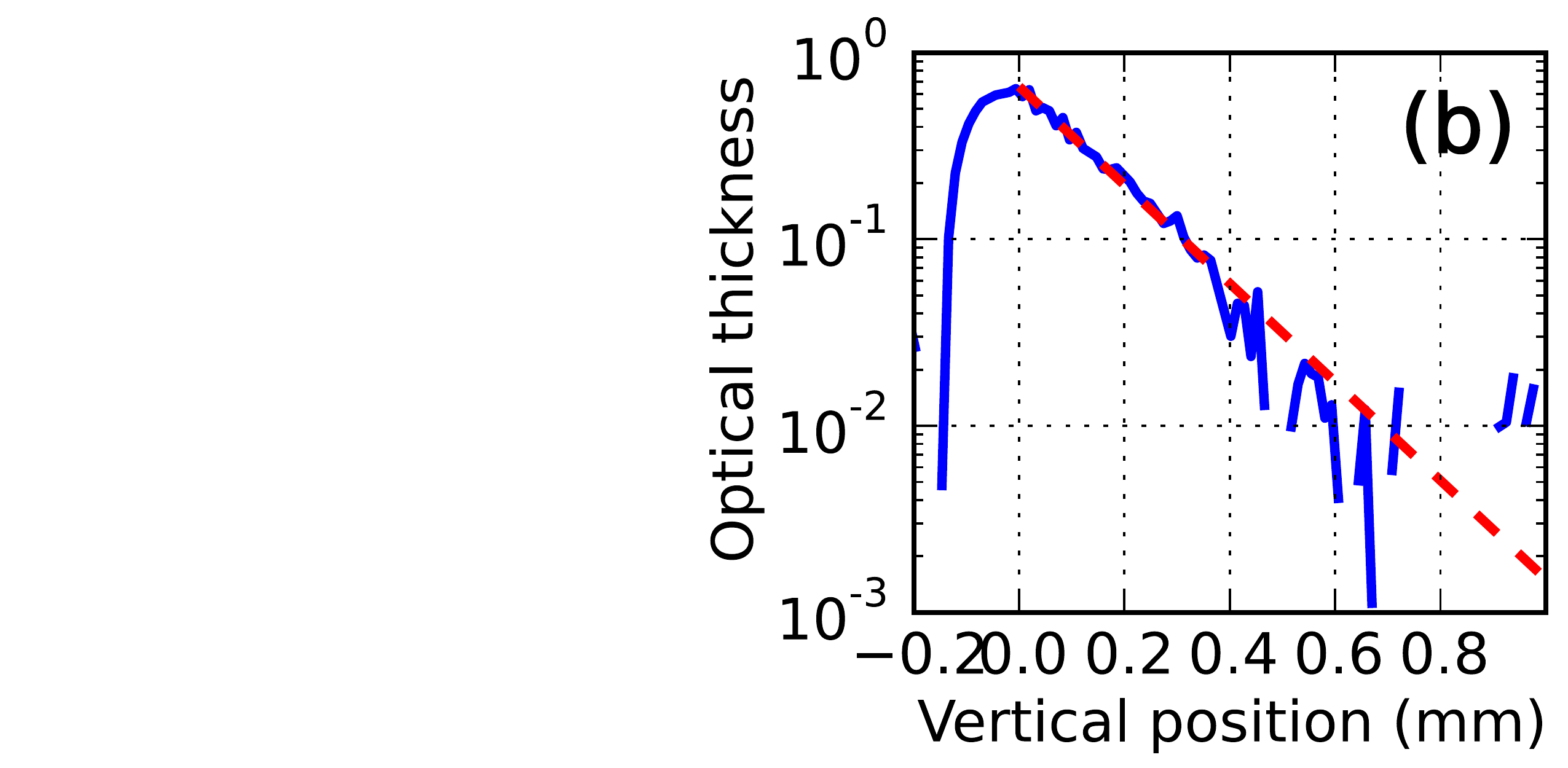}}
\caption{(Color online) (a) \emph{In situ} absorption image of the cloud inside the dipole trap $450 \, \text{ms}$ after loading (sideview). The white dashed line represents the limits of the trap. The atoms lie down on the bottom of the trap due to gravity. (b) Vertical optical thickness profile (blue solid curve). The origin is chosen at the position of the maximum optical thickness. The fit $b(z) = b(0) \exp \left[ -mgz/(k_B T) \right]$ (red dashed line) leads to a temperature of $18 \, \mu\text{K}$. The temperature estimated using the standard time of flight measurement gives $21 \, \mu\text{K}$. Parameters: $P = 200 \, \text{mW}$, $\Delta = 40 \, \text{GHz}$, $L = 1 \, \text{mm}$,  $w = 65 \, \mu \text{m}$, $f_m = 90 \, \text{kHz}$.}
\label{fig7}
\end{figure}

For a trap of size $L$, gravity effects can be neglected when $mgL \ll k_B T$. For a cloud temperature of $40 \, \mu\text{K}$, this condition gives $L \ll 400 \, \mu\text{m}$, showing that gravity is an important parameter for large traps. The influence of gravity can be observed on the absorption imaging from the side of the chamber as shown in Fig. \ref{fig7} where atoms lie down on the bottom of the trap. This situation is similar to the one observed with Strontium MOT on the intercombination line at $689 \, \text{nm}$ \cite{Chaneliere08}.

For atoms at thermal equilibrium in the dipole trap, one expects the optical thickness profile along the vertical direction to be given by the Boltzmann factor
\begin{equation}
b(z) = b(0) \exp  \left[ - \frac{m g z}{ k_B T} \right]. \label{Boltzmann}
\end{equation}
Fig. \ref{fig7} shows the \emph{in situ} density profile of the cloud obtained by absorption imaging. A fit from Eq. (\ref{Boltzmann}) leads to a temperature of $18 \, \mu\text{K}$. The temperature of the cloud given by a time of flight measurement is $21 \, \mu\text{K}$, in good agreement with the \emph{in situ} measurement. As one does not expect a thermodynamic equilibrium to be reached when considering the conservative dynamics of independent particles in a trap, the good agreement between these two methods to estimate the temperature of the atomic cloud suggests that relaxing due to 
residual light scattering combined with a limited trap height or `s-wave' collisions (elastic `s-wave' collision rate after loading $\Gamma_{\text{el}} \lesssim 10 \, \text{s}^{-1}$) might be present in our system.

\subsection{Losses due to `hot' collisions}

The MOT loading time of $\sim 20 \, \text{s}$ allows us to infer a background gas pressure of $P \sim 10^{-9} \, \text{mbar}$ \cite{Arpornthip12}. The trap loss rates due to the background species $i$ can be estimated by \cite{Bjorkholm88, Bali99, Arpornthip12}
\begin{equation}
\gamma_i \simeq 6.8 \frac{P_i}{(k_B T)^{2/3}} \left( \frac{C_i}{m_i} \right)^{1/3} \left( U m_{\text{Rb}}\right)^{-1/6}, \label{LossRate}
\end{equation}
where $T \simeq 300 \, \text{K}$ is the background gas temperature, $P_i$ the partial pressure of the background species $i$ and $U \simeq 190 \, \mu\text{K}$ is the potential height of the dipole trap. With $m_{\text{Rb}}$, $m_i$ we indicate the masses of Rubidium and of the background species $i$ and $C_i$ are the coefficients of the van der Walls interaction potential $- C_i / r^6$ between the ground state trapped Rb atoms and $i$. Table \ref{table1} gives the value of the $C_i$ coefficients for Rb-Rb, Rb-He and Rb-H$_2$ collisions and the corresponding trap loss rates $\gamma_i$ for the parameters of the experiment. The estimated trap lifetimes $1/\gamma_i$ resulting from the background collisions are significantly longer than the measured one, allowing us to conclude that the dominant loss mechanism is not due to collisions with the background gas.

\begin{table}[t!]
\centering
\begin{tabular}{c|c|c}
\hline \hline
 $\qquad$ Species $i$ $\qquad$ & $\qquad$ $C_i$ (a.u.) $\qquad$ &$\qquad$ $\gamma_i$ (s$^{-1}$) $\qquad$  \\
\hline \hline
Rb-Rb &  $4430$ & $0.14$ \\
\hline
Rb-He & $36.2$ & $0.08$  \\
\hline
Rb-H$_2$ & $140$ & $0.16$ \\
\hline \hline
\end{tabular}
	\caption{Coefficients $C_i$ of the van der Walls potential between Rubidium and background species $i$ in Hartree atomic unit (a.u.) = $e^2 a_0^5/(4 \pi \epsilon_0)$, where $e$ is the electron charge, $a_0$ is the Bohr radius and $\epsilon_0$ the vacuum permittivity (data taken from \cite{Bali99}). Trap loss rate $\gamma_i$ computed from Eq. (\ref{LossRate}) with $P_i \simeq 10^{-9} \, \text{mbar}$, $U=190 \, \mu\text{K}$, $T=300 \, \text{K}$. }
	\label{table1}
\end{table}

A further confirmation of the marginality of hot collisions is obtained by performing a lifetime measurement turning on LIAD just after loading the trap. Even though this increases the residual gas pressure of the cell by one order of magnitude $P \sim 10^{-8} \, \text{mbar}$, we notice no difference on the measured lifetime.

\subsection{Influence of the trap size} \label{section_size}

The influence of the size of the trap on its lifetime is an important issue to study before starting the compression of the cloud. After loading, the atoms are kept in the trap for $20 \, \text{ms}$, before linearly reducing the trap size for an additional $20 \, \text{ms}$. The end of this stage is used as the initial condition to measure the lifetime of the cloud. We choose this experimental procedure since it allows an efficient trap loading and prevents non-trapped atoms (because of the too small trap volume compared to the dark MOT volume) to be present in the imaging region when measuring the trapped atom number during the first $40 \, \text{ms}$. The initial conditions are chosen such that the initial atomic density is low enough to prevent density effects in the measurement (small `s-wave' collision rate). Note that this experimental procedure does not maintain the potential height constant, as $U \propto 1/L$ (see Eq. (\ref{maxpotbar})).

\begin{figure}[t]
\centering{\includegraphics[height=4.2cm]{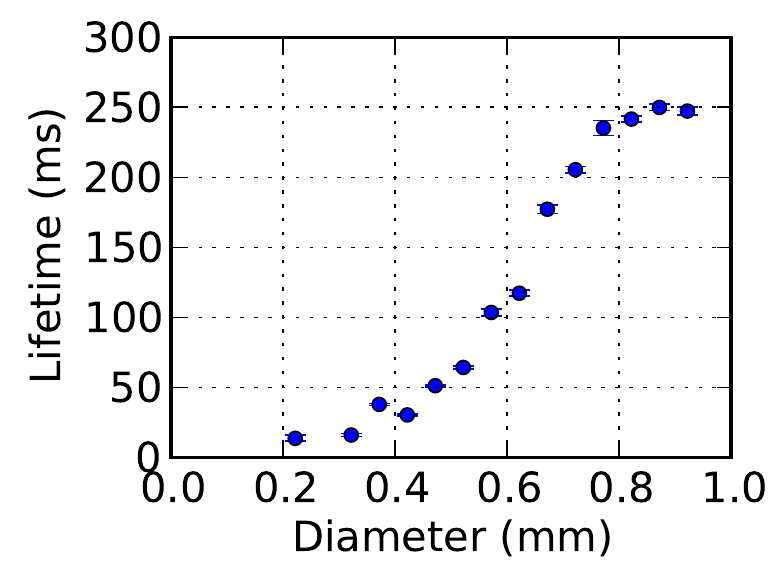}}
\caption{(Color online) Lifetime of the cloud as a function of the dipole trap diameter. Parameters: $P = 200 \, \text{mW}$, $\Delta = 40 \, \text{GHz}$, $w = 65 \, \mu \text{m}$, $f_m = 90 \, \text{kHz}$.}
\label{fig8}
\end{figure}

Fig. \ref{fig8} shows that the lifetime of the cloud rapidly decreases when the size of the trap is reduced. This is consistent with atom losses occurring when the atoms interact with the potential barrier. The smaller the trap is, the more frequent interactions with the barrier are, leading to stronger losses and to a subsequent lifetime reduction. The reduction of the lifetime with the trap size is an important factor to take into account for the cloud compression. 

\subsection{Influence of the rotation frequency} \label{RotFreq}

A first limiting effect gives a lower bound for low rotation frequencies: if during one period $1/f_m$, the atoms with velocity on the order of $\sim \sqrt{k_B T/m}$ move by more than the waist of the laser $w$, the atom can escape the trap between successive arrivals of the laser beam. This condition writes
\begin{equation}
f_m \gg \frac{1}{w} \sqrt{\frac{k_B T}{m}}  \simeq 1 \, \text{kHz},
\end{equation}
for $w=65 \, \mu\text{m}$ and $T = 40 \, \mu\text{K}$. 

\begin{figure}[t]
\centering{\includegraphics[height=4.1cm]{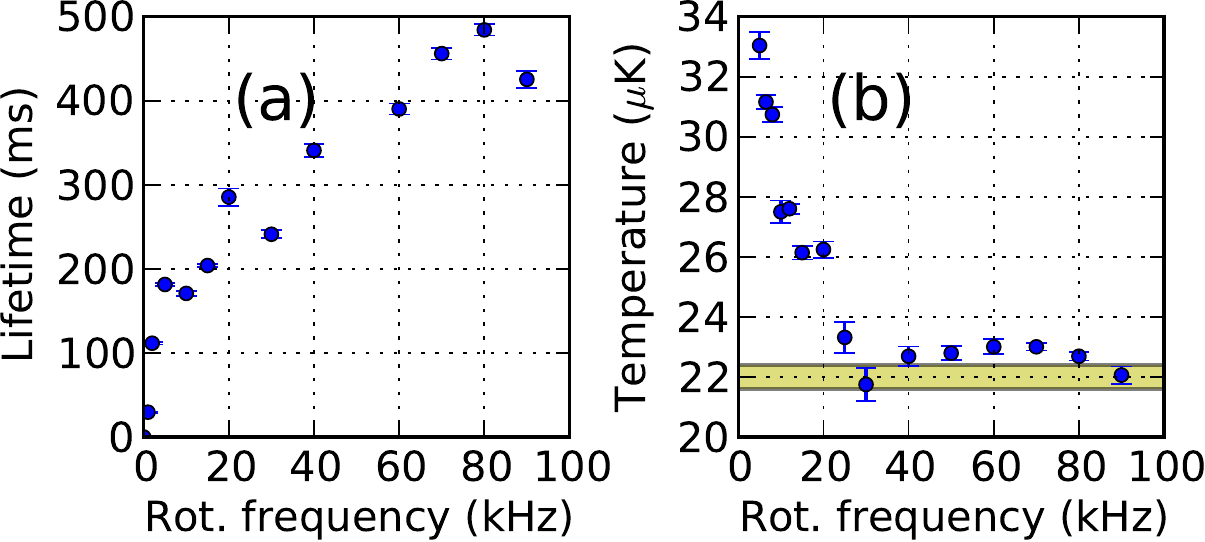}}
\caption{(Color online) (a) Lifetime of the dipole trap as a function of the rotation frequency. (b) Temperature of the cloud measured after $150 \, \text{ms}$ of holding time as a function of the rotation frequency. The shaded area represents the initial dark MOT temperature before loading the dipole trap. Parameters: $P = 200 \, \text{mW}$, $\Delta = 40 \, \text{GHz}$, $L = 1 \, \text{mm}$, $w = 65 \, \mu \text{m}$.}
\label{fig9}
\end{figure}

In order to study the influence of the rotation frequency above this limit, we vary the rotation frequency and keep all other parameters constant. Fig. \ref{fig9} (a) shows that the lifetime increases with increasing rotation frequency without reaching a saturation in the explored frequency range (contrary to what is observed in \cite{Davidson00}). This can be understood by noticing that the faster the beam rotates, the better the timed-averages potential approximation is. The upper bound for the rotation frequency is set by the input modulation bandwidth of the VCOs.

Fig. \ref{fig9} (b) shows the cloud temperature after $150 \, \text{ms}$ of holding time as a function of the rotation frequency. Below $30 \, \text{kHz}$, we observe an important heating which decreases with increasing rotation frequency. Above $30 \, \text{kHz}$, the temperature reaches a plateau which is close to the initial dark MOT temperature (shaded area on Fig. \ref{fig9} (b)). We identify this phenomenon as being due to dipolar heating, i.e. heating due to the dipole potential fluctuations that atoms experience when they bounce on the potential barrier because of the rotating laser. From this measurement, we can get an estimate of the dipole induce heating rate, expected to scale as $1/(L f_m)$, which can be used for optimizing compression schemes.

After compression, for small trap size in the harmonic regime, we also require the rotation frequency to be large compared to the trap frequency $\omega/(2 \pi)$ for the timed-averaged approximation to be valid. The maximum trap frequencies that can be reached are a few kHz. Using a rotation frequency close to $100 \, \text{kHz}$ the two previous conditions are easily satisfied.

It is clear that the trap has better performances when the rotation frequency is increased. However, there is  an intrinsic limitation that restrains this argument. Indeed modulating a signal at a frequency $f_m$, creates sidebands separated by $f_m$. These sidebands affect the dipole potential. The limit for having a nice `continuous trap shape' is to consider that the distance $d$ between two peaks associated to two sidebands should be smaller than the laser waist $w = 65 \, \mu\text{m}$. The AOM deflection angle is  $\alpha = 9.3 \times10^{-5} \, \text{rad MHz}^{-1}$, and after a distance $L = 150 \, \text{mm}$ (the lens focal length), the distance between two points is $d = \alpha f_m L$. The condition then writes
\begin{equation}
f_m \ll \frac{w}{\alpha L} \simeq 5 \, \text{MHz}.
\end{equation}
This argument shows that there is an intrinsic limit for the maximum frequency one can use. However, we are limited by $f_m < 100 \; \text{kHz}$ by the VCOs, so there is still room for improvement.

In summary, below a modulation frequency of $\sim 1 \, \text{kHz}$, the atoms are no longer trapped. This regime can be observed on the two first points of Fig. \ref{fig9} (b). In the intermediate regime $3 \, \text{kHz} < f_m < 30 \, \text{kHz}$, we observe a reduction of the heating. This regime is associated to dipolar heating losses due to the rotating laser. For $f_m > 30 \, \text{kHz}$ we observe an increase of the lifetime and a stabilization of the temperature.
These results justify the choice of a rotation frequency of $90 \, \text{kHz}$, which corresponds to the maximum frequency we can use taking into account the $100 \, \text{kHz}$ input modulation bandwidth of the VCOs. The performances of the trap might probably be improved using VCOs with a higher input modulation bandwidth and a better frequency stability.

\subsection{Influence of the detuning}

\begin{figure}[t]
\centering{\includegraphics[height=4.1cm]{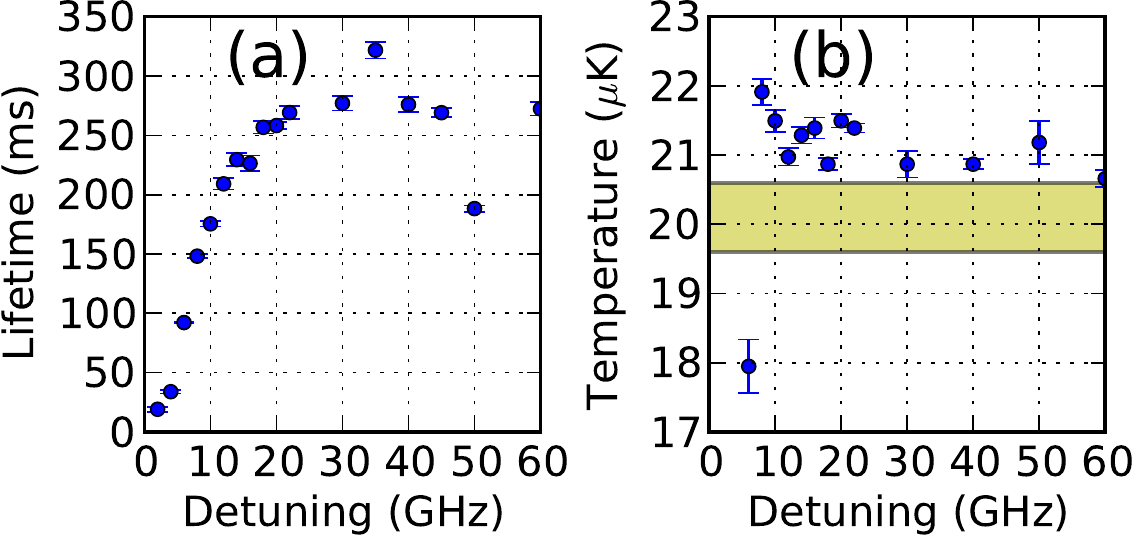}}
\caption{(Color online) (a) Lifetime of the dipole trap as a function of the laser detuning at constant potential height. For this measurement we adjust the laser power to maintain a constant potential barrier $U \propto P/\Delta$ while the detuning is varied. (b) Temperature of the cloud measured after $150 \, \text{ms}$ of holding time as a function of the laser detuning. The shaded area represents the initial dark MOT temperature before loading the dipole trap. Parameters: $L = 1 \, \text{mm}$,  $w = 65 \, \mu \text{m}$, $f_m = 90 \, \text{kHz}$.}
\label{fig10}
\end{figure}

We study the influence of the detuning on the lifetime and temperature by changing $\Delta$ and $P$ maintaining $U \propto P/\Delta$ constant. The first point on the right side of Fig. \ref{fig10} is measured for $P = 200 \, \text{mW}$ and $\Delta = 60 \, \text{GHz}$, corresponding to a potential height $U = 126 \, \mu\text{K}$.

Fig. \ref{fig10} (a) shows that for $\Delta < 20 \, \text{GHz}$ the lifetime strongly depends on the detuning, and that for $\Delta > 20 \, \text{GHz}$, the lifetime becomes constant $\sim 280 \, \text{ms}$. It is important to note that the semiconductor laser diodes used in these experiments present, in addition to their main laser mode, a pedestal which spreads over $40 \, \text{nm}$. From the data of Fig. \ref{fig10} (a), we conclude that above $20 \, \text{GHz}$, spontaneous emission due to the main laser mode is not the limiting phenomenon leading to atom losses. We attribute the limited lifetime above $20 \, \text{GHz}$ to heating induced by the amplified spontaneous emission pedestal and dipolar heating as discussed in section \ref{RotFreq}.

On Fig. \ref{fig10} (b), we notice that the cloud temperature does not depend on the laser detuning. The temperature is constant, slightly higher than the dark MOT temperature. After loading the cloud at the dark MOT temperature spontaneous emission leads to atom losses without temperature increase (the potential height is $126 \, \mu\text{K}$). We conclude that the extra heating due to spontaneous emission is immediately suppressed by evaporation: the atoms escape from the trap when their total energy becomes higher than the potential barrier height.


\begin{figure}[t]
\centering{\includegraphics[height=4.2cm]{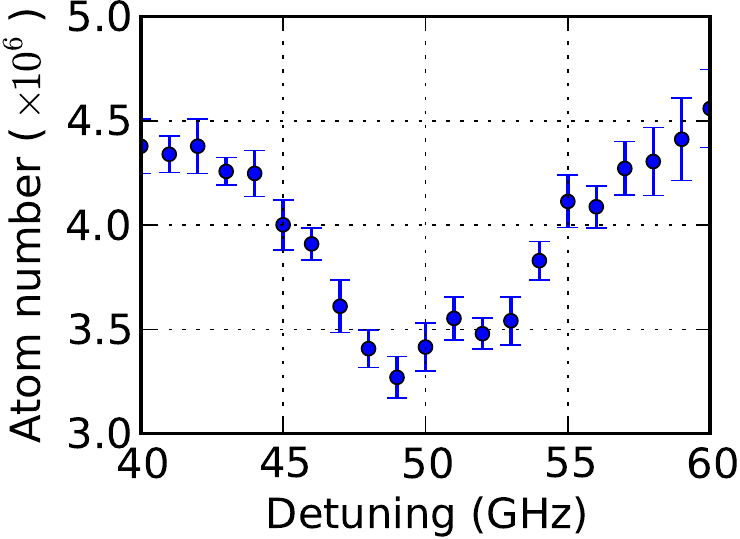}}
\caption{(Color online) Number of atoms in the dipole trap after $150 \, \text{ms}$ of holding time as a function of the laser detuning at constant potential height. This curve shows a region where we measure abnormally low lifetimes which indicates a resonance whose origin is unknown. The resonance is centered around $50 \, \text{GHz}$ and is $10 \, \text{GHz}$ broad. Parameters: $L = 1 \, \text{mm}$, $w = 65 \, \mu \text{m}$, $f_m = 90 \, \text{kHz}$ (same as those of Fig. \ref{fig10}).} 
\label{fig11}
\end{figure}

We notice on Fig. \ref{fig10} (a), particularly short trap lifetimes around $\Delta = 50 \, \text{GHz}$. In order to study this more quantitatively, we measure the number of atoms in the dipole trap after $150 \, \text{ms}$ of holding time using the same experimental parameters as in Fig. \ref{fig10}. These data points are shown in Fig. \ref{fig11} where we observe a $10 \, \text{GHz}$ broad resonance centered around $50 \, \text{GHz}$ which manifests itself as atom losses. No resonance is seen in the incident laser spectrum and its origin remains elusive. Possible explanations might involve acoustic modes in the optical fiber or more interestingly molecular resonances.

\subsection{Influence of the laser system spectrum}

\begin{figure}[t]
\centering{\includegraphics[height=8.4cm]{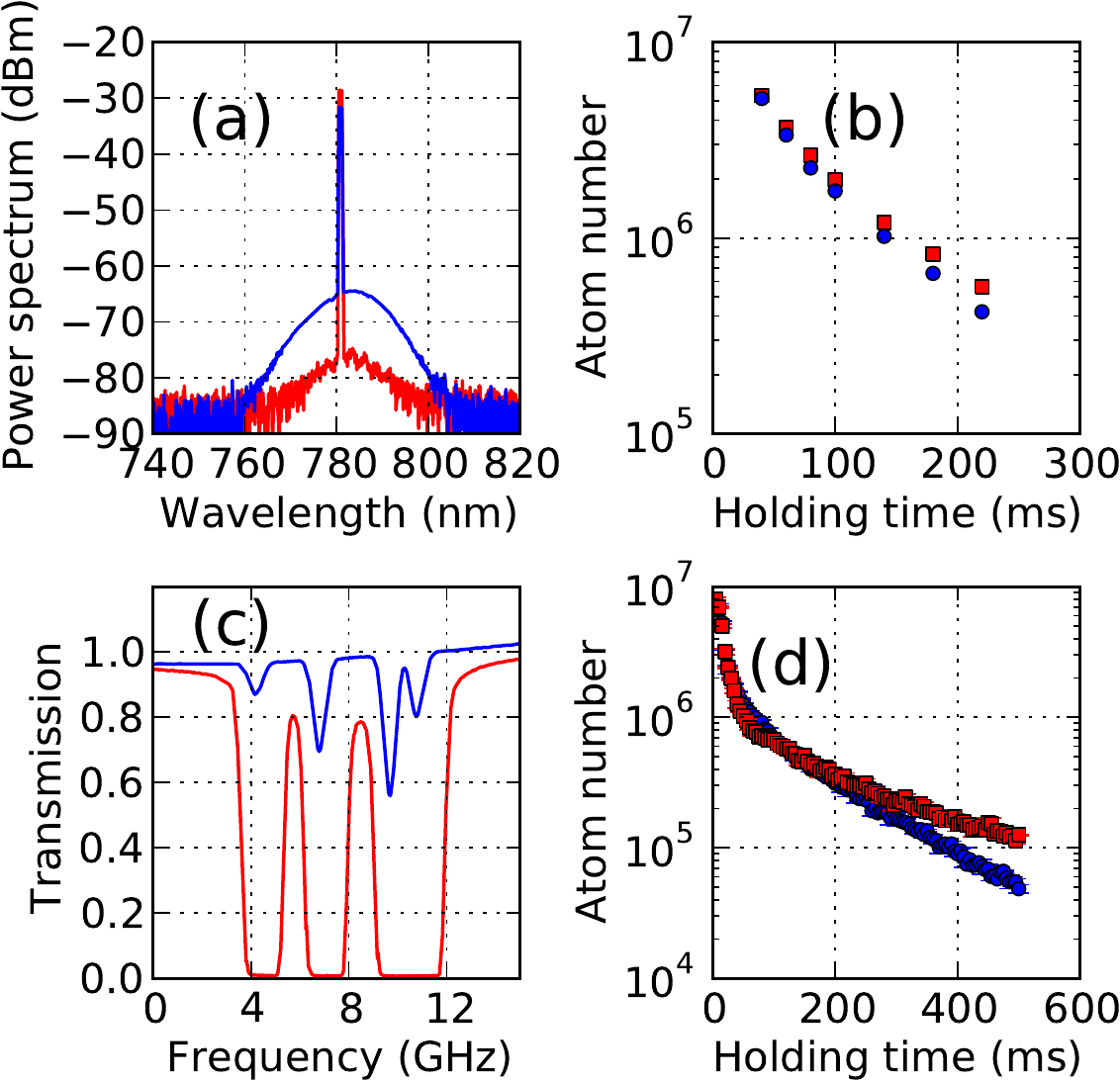}}
\caption{(Color online) (a) Power spectrum of the laser system (blue curve) and of the laser system plus etalon (red curve). (b) Number of atoms in the dipole trap as a function of the holding time without (blue points) and with etalon (red squares). The lifetimes are $58 \, \text{ms}$ and $66 \, \text{ms}$ respectively. The laser power is the same for the two measurements. Parameters: $P = 70 \, \text{mW}$, $\Delta = 24 \, \text{GHz}$, $L = 1 \, \text{mm}$, $w = 65 \, \mu \text{m}$, $f_m = 90 \, \text{kHz}$. (c) Transmission of a weakly saturating probe through a $7.5 \, \text{cm}$ long Rubidium cell at $22 \, ^\circ\text{C}$ (blue curve) and $120 \, ^\circ\text{C}$ (red curve). (d) Number of atoms in the dipole trap as a function of the holding time using the Rubidium cell at $22 \, ^\circ\text{C}$ (blue points) and $120 \, ^\circ\text{C}$ (red squares). Parameters: $P = 200 \, \text{mW}$, $\Delta = 43 \, \text{GHz}$, $L = 1 \, \text{mm}$, $w = 65 \, \mu \text{m}$, $f_m = 90 \, \text{kHz}$.} 
\label{fig12}
\end{figure}

Assuming thermal equilibrium, the mean photon scattering rate $\Gamma_{\text{sc}}$ can be estimated from the temperature by computing the average potential
\begin{equation}
\langle U \rangle = \frac{\int \mathrm d \mathbf r \, U(\mathbf r) \exp \left[ - \frac{U(\mathbf r) + mgz}{k_B T} \right] }{ \int \mathrm d \mathbf r \, \exp \left[ - \frac{U(\mathbf r) + mgz}{k_B T} \right] }, \label{avg_pot}
\end{equation}
and using the relation $\hbar \Gamma_{\text{sc}} = U \, \Gamma / \Delta$. For blue dipole trap $\langle U \rangle$ and $\langle \Gamma_{\text{sc}} \rangle$ are both increasing functions of the temperature. For the typical parameters of the experiment $P = 200 \, \text{mW}$, $\Delta = 40 \, \text{GHz}$, $L = 1 \, \text{mm}$,  $w = 65 \, \mu \text{m}$, $T=20 \, \mu\text{K}$, evaluating Eq. (\ref{avg_pot}) gives $\langle U \rangle = 2.11 \, \mu\text{K}$ and $\langle \Gamma_{\text{sc}} \rangle = 41 \, \text{s}^{-1}$. This underestimates the photon scattering rate by not accounting for imperfections of the potential. More importantly, a major drawback of semiconductor laser systems is their amplified spontaneous emission background which spreads over $40 \, \text{nm}$ and contains photons resonant with the atomic lines. These photons contribute to heating, leading to a potential reduction of the trap lifetime. The spontaneous emission background is clearly seen when looking at the laser system power spectrum shown in Fig. (\ref{fig12}) (a). It represents $0.9 \, \%$ of the total laser power.  In this section, we investigate two different approaches to filter resonant photons using an etalon or a Rubidium cell.

The first method consists in filtering the spontaneous emission background using an etalon of finesse $\mathcal F = 60$ with free spectral range $\text{FSR} = 210 \, \text{GHz}$. Its transmission exhibits peaks separated by $210 \, \text{GHz}$ with full with at half maximum $\text{FSR}/\mathcal F = 3.5 \, \text{GHz}$. For an optical spectrum analyzer that does not resolve the transmission peaks, the expected reduction of the spontaneous emission background is $10 \log_{10} \mathcal F  = 17 \, \text{dB}$, which shows good agreement with the measurements presented in Fig. (\ref{fig12}) (a). After filtering, the spontaneous emission background only represents $0.04 \, \%$ of the total laser power. Due to the poor mode quality of the laser system, the coupling efficiency through the etalon is weak letting only $70 \, \text{mW}$ of light available for creating the dipole trap. In order to maintain a decent trap depth, we reduce the laser detuning to $24 \, \, \text{GHz}$. The influence of the filtering on the trap lifetime is shown in Fig. (\ref{fig12}) (b). The measurement indicates that the spontaneous emission background is not the dominant effect which limits the trap lifetime below $200 \, \text{ms}$.

The second method involves using a $7.5 \, \text{cm}$ long Rubidium cell to filter photons around the Doppler broadened lines. Heating the cell from $22 \, ^\circ\text{C}$ to $120 \, ^\circ\text{C}$ allows filtering photons by increasing the Rubidium pressure by several orders of magnitude. Using heated cells as narrow band absorption filters to reduce the amplified spontaneous emission background of diode laser systems has proven to be an efficient technique to minimize resonant photon scattering in dipole traps; hence extending their lifetimes \cite{Hammes01,Dumke02,Lignier05}. Fig. \ref{fig12} (c) illustrates the filtering efficiency by looking at the transmission of a weakly saturating probe through the cell. Fig. \ref{fig12} (d) shows the number of atoms in the trap as a function of time with and without filtering. These curves show a double exponential decay. We relate the first decay to losses occurring during the transient regime and the second one to losses in the quasi-steady regime where the lifetime of the trap is evaluated. Again, below $200 \, \text{ms}$ we are not able to observe any benefit from resonant photon filtering. However, for longer trapping times, we notice less atom losses indicating that photon scattering from the spontaneous emission background becomes important.

In conclusion, below $200 \, \text{ms}$, light scattering from the laser pedestal does not limit the experiment. Since we are aiming at fast compression times, we choose not to use any filtering method in the setup. Nevertheless, our data indicates that one should definitely consider implementing these techniques to obtain long trapping times which might be useful in applications involving quantum degenerate gases.

\subsection{Conclusion and final performances}

\begin{table*}[t!]
\centering
\begin{tabular}{p{0.32\linewidth}|p{0.32\linewidth}|p{0.32\linewidth}}
\hline \hline
 Experimental parameters & Losses/heating mechanisms & Solutions   \\
\hline \hline
Size $L$ &  Interactions with potential barrier  & Largest possible size  $\sim 1 \, \text{mm}$ for better lifetime/loading\\
\hline
Rotation frequency $f_{\text{m}}$ & Dipolar heating & $f_{\text{m}} > 30 \, \text{kHz}$, largest possible keeping in mind the sideband limit. Limited by VCOs modulation bandwidth: $90 \, \text{kHz}$ \\
\hline
Laser mode detuning $\Delta$ & Spontaneous emission & $\Delta > 20 \, \text{GHz}$, ideally the largest possible while keeping a suitable potential height \\
\hline
Laser system pedestal & Spontaneous emission & Beam filtering through etalon/Rubidium hot cell\\
\hline \hline
\end{tabular}
	\caption{Summary of the experiments performed on the static dipole trap in the \emph{quasi-steady regime}. For each measurement, we identify a mechanism leading to losses/heating and find the parameters that give the best trap performances.}
	\label{table2}
\end{table*}

The dipole trap presented in this section allows for trapping a large number of atoms with an excellent loading efficiency. Table \ref{table2} summarizes the performances and limitations of the setup. It helps us find the best compromise for our purposes with the following parameters that maximize the loading, the lifetime and the cloud temperature: $P = 200 \, \text{mW}$, $\Delta = 40 \, \text{GHz}$, $L = 1 \, \text{mm}$, $w = 65 \, \mu\text{m}$, $f_m = 90 \, \text{kHz}$. With these parameters, the height of the potential barrier is $\sim 190 \, \mu\text{K}$, the typical lifetime is $\sim 450 \, \text{ms}$ and the temperature of the cloud is constant $\sim 25 \, \mu\text{K}$ (about the dark MOT temperature) leading to a factor $\eta = U/(k_B T) \simeq 7.6$.

\section{Compression}

In this section, we compress the atomic cloud by dynamically reducing its size $L$ by means of the blue detuned crossed dipole trap. The goal is to quickly compress the maximum number of atoms $N$ below the strong localization threshold, qualitatively given by the Ioffe-Reggel criterion \cite{Ioffe60}. As discussed in the introduction, this threshold corresponds to atomic densities of $n= N/L^3 \simeq10^{13} - 10^{14} \, \text{cm}^{-3}$ for Rubidium atoms.

\subsection{Principles of the compression scheme} \label{PremThoughts}

The aim of this section is to understand the relevant parameters for compressing the cloud. Simple arguments allow to qualitatively address important issues, even though they do not aim at describing the experiment in a rigorous way (in particular concerning the role of gravity). 

\subsubsection*{Maximum compression speed} \label{maxcompspeed}

We are interested in calculating the maximum speed the potential barrier of height $U$ can move before the atoms are no longer able to follow the motion of the barrier and consequently `jump' over it. Let us consider the most pessimistic case where the potential barrier moves towards the atoms in the laboratory frame at a speed $v_{\text{pot}}$ while the atom moves in the opposite direction at a speed $v_{\text{atom}}$. In the frame attached to the potential barrier, the atom has a velocity $v = v_{\text{pot}} + v_{\text{atom}}$ and will not `jump' over the barrier if its kinetic energy is smaller than the potential height leading to the criterion $v_{\text{pot}} < \sqrt{2U/m}-v_{\text{atom}}$. Noting that $\sqrt{2U/m} \gg v_{\text{atom}}$, the condition simplifies to
\begin{equation}
v_{\text{pot}} < \sqrt{\frac{2U}{m}}. \label{speed_criterion}
\end{equation}
For a $200 \, \mu\text{K}$ potential height, we obtain a maximum barrier velocity of $0.2 \, \text{m} \, \text{s}^{-1}$. If the barrier moves by $500 \, \mu\text{m}$, the minimal time to compress the cloud is $2.5 \, \text{ms}$. This will not be a restrictive constraint for the experimental realization of the compression. Moreover, knowing that the potential barrier increases during compression (cf. Eq. (\ref{maxpotbar})), Eq. (\ref{speed_criterion}) overestimates the minimum compression time.

\subsubsection*{Heating}

For an ideal gas undergoing an adiabatic reversible process the following equation applies $TV^{\gamma-1} = \text{constant}$, where $T$ is the temperature, $V$ the volume and $\gamma$ the adiabatic index. For a monatomic gas $\gamma = 5/3$. When the size of the trap is reduced to $L_f$, a cloud of monatomic atoms (e.g. Rubidium atoms) initially at a temperature $T_i$ in a trap of size $L_i$, will reach a temperature $T_f$ given by
\begin{equation}
\frac{T_f}{T_i} = \left( \frac{L_i}{L_f} \right)^2. \label{TempInc}
\end{equation}
Any non adiabatic reversible compression would lead to higher final temperatures than the one predicted by Eq. (\ref{TempInc}).

\subsubsection*{Phase space density evolution}

When the size of the trap is reduced such that $a \simeq w$ i.e. the radius is about equal to the waist, the geometry of the trap changes from a box with Gaussian walls to a harmonic potential $U = U_1 + 1/2 \times m \omega^2 r^2$ where the analytical expressions for $U_1$ and $\omega$ are given in section \ref{TrapConf}. During an adiabatic compression, the entropy of the cloud is conserved implying phase space density conservation when the geometry of the trap does not change. However, when the geometry of the trap changes from a box to a harmonic potential, the phase space density slightly increases. Noting $\text{PSD}_i \equiv n_i \Lambda_{T_i}^3$ ($n_i$ the density and $\Lambda_{T_i}$ the initial thermal wavelength) the initial phase space density for the atom in a box, and $\text{PSD}_f \equiv n_f \Lambda_{T_f}^3$ the final phase space density for the atoms at the center of the harmonic trap, one can easily show that the increase in phase space density for an adiabatic transformation is given by \footnote{The following equation is obtain by equating the entropy of an ideal gas in a box $S = N k_B \left\{ \ln \left[ \frac{V}{N \Lambda_T^3} \right] + \frac{5}{2} \right\}$ and its entropy in a harmonic trap $S = N k_B \left\{ \ln \left[ \frac{1}{N} \left( \frac{k_B T}{\hbar \omega} \right)^3 \right] + 4  \right\}$.}
\begin{equation}
\text{PSD}_f = e^{3/2} \text{PSD}_i.
\end{equation}
This phase space density increase comes from the modification of the density of states when the trap geometry is modified \cite{Pinkse97, Stamper98}.

\subsubsection*{Potential height evolution}

The temperature increase of the cloud resulting from the compression tends to make the atoms escape from the trap by `jumping' over the potential barrier. However, when the trap size is reduced, the potential height increases as well. Eq. (\ref{maxpotbar}) shows that a trap with initial size $L_i$, potential height $U_i$, compressed to a final size $L_f$, has a final potential height $U_f = U_i L_i/L_f$. The temperature of the atoms thus increases faster than the height of the potential. Using Eq. (\ref{TempInc}), and neglecting collisions (evaporation), atoms stay in the trap while $k_B T_f = U_f$, which allows to estimate the size of the trap before the atoms escape
\begin{equation}
\frac{L_f}{L_i} =  \frac{k_B T_i}{U_i}.
\end{equation}
In the experiment $k_B T_i/U_i \sim 1/8$ so that starting from a trap size of $1 \, \text{mm}$, we can compress the cloud down to $\sim 125 \, \mu\text{m}$ before atoms `jump' over the potential barrier.

\subsubsection*{Lifetime constraint}

Besides the constraints on the minimum compression time discussed in section \ref{maxcompspeed}, there are also restrictions about the maximum compression time. We have shown in section \ref{section_size} that the trap lifetime strongly depends on its size. This implies important limitations on the maximum compression time that should be used. For example, using the data from Fig. \ref{fig8}, compressing the cloud to a final size of $400 \, \mu\text{m}$ should be done in less than $30 \, \text{ms}$ (which is the lifetime of the trap for this diameter) in order not to loose too many atoms during compression.

\subsection{Experimental realization} \label{ExpRea}

\begin{figure}[t]
\centering{\includegraphics[height=4.2cm]{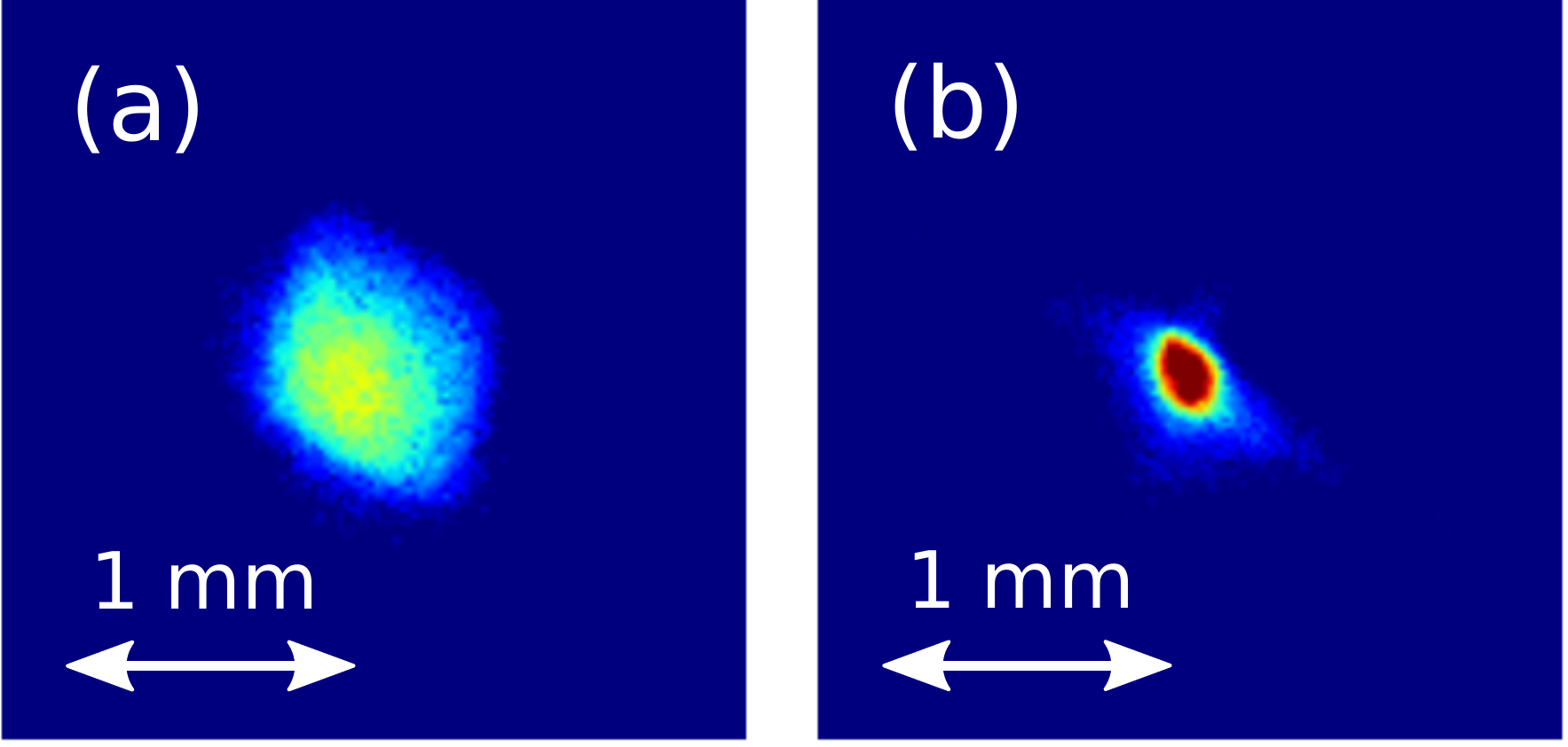}}
\caption{(Color online) \emph{In situ} fluorescence images of the atoms in the dipole trap taken from the top of the cell. (a) Cloud before compression, trap size is $L_i = 1 \, \text{mm}$. (b) Cloud after a $10 \, \text{ms}$ compression, trap size is $L_f = 300 \, \mu\text{m}$. Parameters: $P = 200 \, \text{mW}$, $\Delta = 40 \, \text{GHz}$,  $w = 65 \, \mu \text{m}$, $f_m = 90 \, \text{kHz}$.}
\label{fig13}
\end{figure}

After loading the dipole trap (initial size $L_i = 1 \, \text{mm}$) using the protocol described in section \ref{paramexpseq}, the size of the trap is kept constant during $20 \, \text{ms}$ in order for the non-trapped atoms to escape from the imaging field of view. The size of the trap is then linearly reduced and the compression time can be varied. The data presented below are taken for different final trap sizes $L_f$, while the compression time is fixed. Fig. \ref{fig13} shows fluorescence images of the cloud viewed from the top of the cell, before and after compression. Absorption imaging from the side of the trap is used to perform quantitative measurements.

\begin{figure}[t]
\centering{\includegraphics[height=8.4cm]{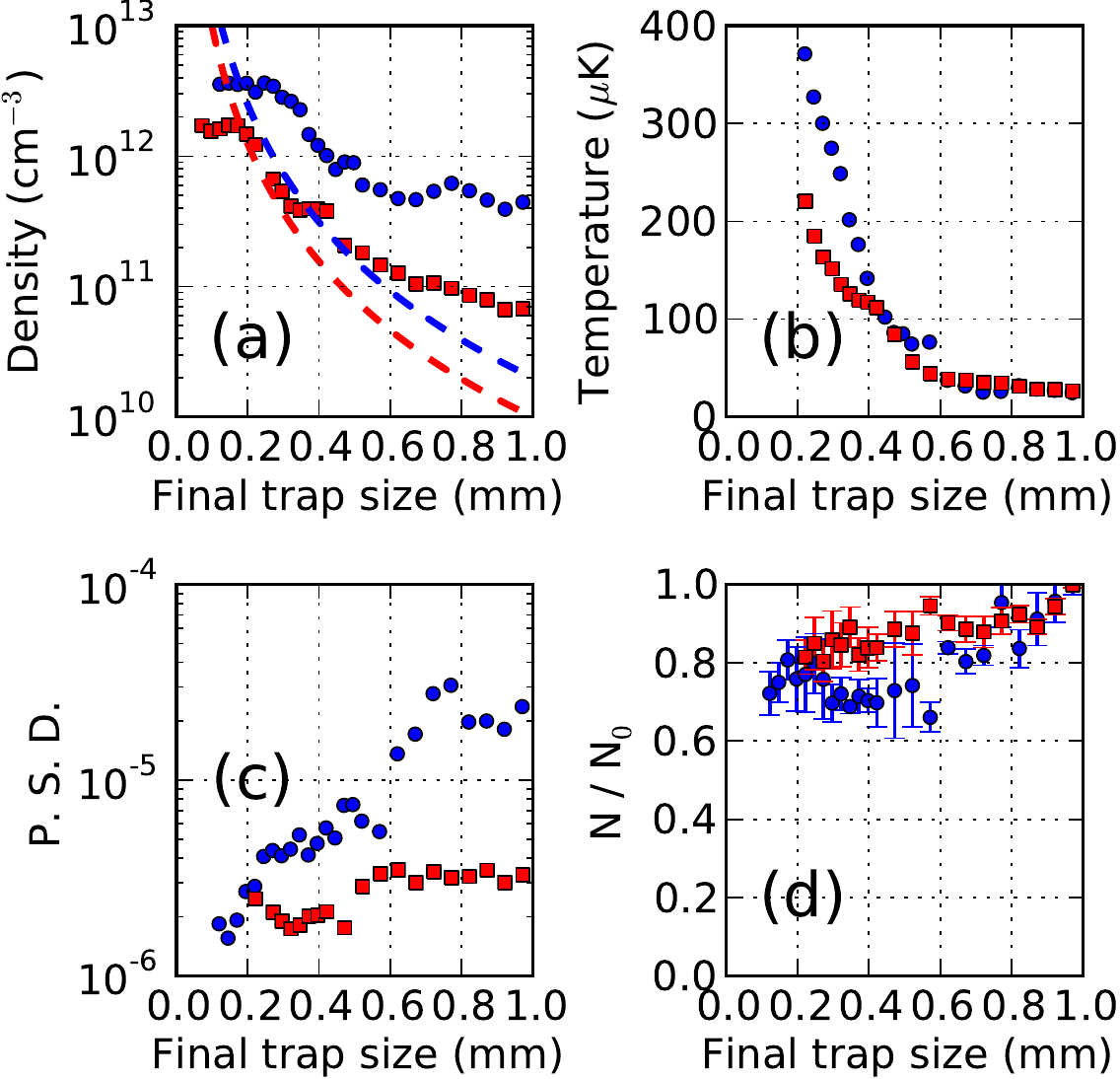}}
\caption{(Color online) (a) Density, (b) temperature, (c) phase space density of the cloud as a function of the trap size after compression $L_f$. (d) Fraction of the remaining atoms after compression. The initial trap size is $L_i = 1 \, \text{mm}$. The blue points correspond to a $5 \, \text{ms}$ compression and the red squares correspond to a $10 \, \text{ms}$ compression. The initial conditions are $N=2 \times 10^7$ initially loaded atoms for the blue points and $N=10^7$ atoms for the red squares. The dashed curves in (a) corresponds to the densities one would obtain if the trap was uniformly loaded $n = N/L^3$. Parameters: $P = 200 \, \text{mW}$, $\Delta = 40 \, \text{GHz}$, $w = 65 \, \mu \text{m}$, $f_m = 90 \, \text{kHz}$.}
\label{fig14}
\end{figure}

Fig. \ref{fig14} shows the density (a) and the temperature (b) of the cloud as a function of the trap size after compression $L_f$ for two different compression times: $5 \, \text{ms}$ (blue points) and $10 \, \text{ms}$ (red squares). The blue points correspond to $2\times10^7$ initially loaded atoms and the red squares to $10^7$ atoms. We manage to increase the density by more than one order of magnitude to reach $5 \times 10^{12} \, \text{cm}^{-3}$ reducing the trap size by a factor of $5$. Fig. \ref{fig14} (b) shows strong heating during compression. The faster the compression, the higher the heating, as expected from the discussion on adiabatic heating in section \ref{PremThoughts}. In the future, if the high final temperature is a problem, one can think of adding an evaporative cooling stage after compression which would make the cloud colder and denser. We notice that for small compression ($L_f > 0.6 \, \text{mm}$), the density and the temperature are almost not affected. This can be understood by looking at absorption imaging of the cloud (see Fig. \ref{fig7}) where the atoms do not occupy the full trap volume because of gravity. The change of slope that we observe for $L_f < 0.6 \, \text{mm}$ corresponds to the situation where the cloud starts occupying all the trap volume. For small trap size we observe a saturation of the cloud density when the trap reaches the harmonic regime. This might be due to the the increased sensitivity to instabilities for small traps. Indeed, in the harmonic regime, the trap frequency is very sensitive to the trap size. Inelastic (`s-wave' or/and light assisted) collisions can also play a role at high densities and contribute to the observed fluctuations.

Fig. \ref{fig14} (c) shows the phase space density $n \Lambda_T^3$ as a function of the final trap size. As previously discussed, the phase space density is conserved for an adiabatic compression if the trap geometry does not change (the density of states is conserved). We observe that for a $10 \, \text{ms}$ compression time, phase space density is conserved but it is no longer the case when the compression time is $5 \, \text{ms}$.

Fig. \ref{fig14} (d) shows the fraction of remaining atoms after compression. We manage to keep more than $80 \, \%$ of the atoms which is very promising. To do so, we need to compress faster than any loss mechanism. In particular, faster that the trap lifetime for the final trap size we are aiming at (see Fig. \ref{fig8}).

\subsubsection*{Maximum density}

The experiments performed above are designed to understand how compression works. They are carried out using a $2.5 \, \text{s}$ MOT loading time from a low pressure background gas resulting in a modest number of atoms loaded into the trap $\sim 10^7$. Techniques to increase the number of trapped atoms include making the loading time longer, rising the background gas pressure using LIAD or enlarging the initial trap volume. The latter is something difficult to achieve with our current setup. Therefore, a cloud with more atoms but a larger initial size would not increase significantly the number of trapped atoms. Improving the dark MOT density (by further detuned trapping lasers) will be important to increase the number of trapped atoms well beyond $10^8$.

To test the performances of the trap we simply increase the MOT loading time to $20 \, \text{s}$. We also use a detuning of $20 \, \text{GHz}$ (compared to $40 \, \text{GHz}$ usually) to increase the potential height. In these conditions, $5\times10^7$ atoms are loaded into the trap. After compressing the cloud to a final size of $200 \, \mu\text{m}$ in $5 \, \text{ms}$, we measure a density of $10^{13} \, \text{cm}^{-3}$, which corresponds to
\begin{equation}
k \cdot l \simeq 2.8,
\end{equation}
compatible with the Ioffe-Reggel criterion. This setup thus proves to be an effective tool to rapidly compress a large atomic cloud down to the strong localization threshold paving the way for efficient exploration of light-matter interaction in the dense regime. In the future, the trap performances can be further improved by having a more precise control on the trap size evolution when the trap enters the harmonic regime since in this regime the parameters of the trap (e.g. the trap frequency) vary very fast with its radius.

\subsection{Collisions and thermalization}

\begin{figure}[t]
\centering{\includegraphics[height=4.2cm]{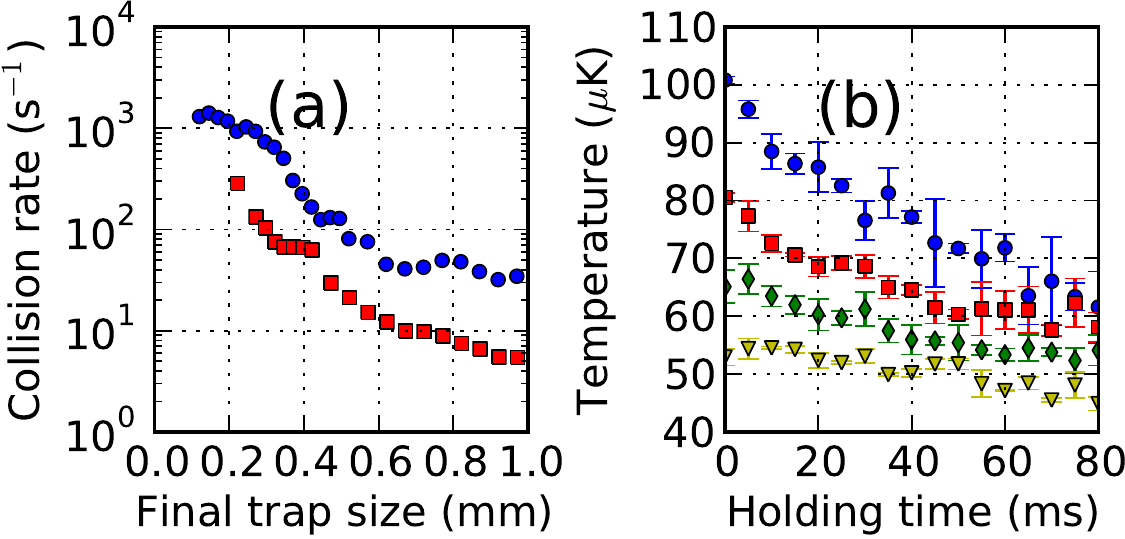}}
\caption{(Color online) (a) Elastic `s-wave' collision rate as a function of the trap size after compression. The blue points correspond to a $5 \, \text{ms}$ compression and $2 \times 10^7$ atoms initially loaded. The red squares correspond to a $10 \, \text{ms}$ compression and $10^7$ atoms initially loaded. (b) Temperature as a function of the holding time for different initial conditions of the atoms in the dipole trap. The different initial conditions are prepared by compressing the cloud during $40 \, \text{ms}$ from an initial size $L_i = 1 \, \text{mm}$ to different final trap sizes after compression $L_f = 0.42 \, \text{mm}$ (blue points), $0.47  \, \text{mm}$ (red squares), $0.52  \, \text{mm}$ (green diamonds), $0.57  \, \text{mm}$ (yellow triangles) respectively. After compression the diameter is kept constant. The state of the system after compression is used as the initial condition for this measurement. The smaller the trap size after compression is, the hotter and denser the cloud initial conditions are. The initial densities and temperatures are respectively of $1.7\times10^{11}, 1.6\times10^{11}, 1.4\times10^{11}, 1.2\times10^{11} \, \text{cm}^{-3}$ and $101, 81, 65, 53 \, \mu\text{K}$ which correspond to elastic `s-wave' collision rates of $27, 23, 17, 14 \, \text{s}^{-1}$ respectively. These `s-wave' collision rates are compatible with the points showing the temperature reduction. Parameters: $P = 200 \, \text{mW}$, $\Delta = 40 \, \text{GHz}$, $w = 65 \, \mu \text{m}$, $f_m = 90 \, \text{kHz}$.}
\label{fig15}
\end{figure}

When the cloud is compressed, the spatial density and the temperature rise, which makes the cloud enter into a regime where collisions are no longer negligible on the timescale of the experiments \cite{Ketterle96}. The elastic collision rate is given by
\begin{equation}
\Gamma_{\text{el}} = n \sigma \bar v_{\text{rel}}, \label{collrate}
\end{equation}
where $n$ is the cloud density, $\bar v_{\text{rel}} = 4 \sqrt{k_B T / (\pi m)}$ is the mean relative atom velocity and $\sigma$ is the total elastic cross section. If we consider pure `s-wave' collisions, which in our case is a strong approximation given the temperature of the gas (where higher order collisions, e.g. `p-wave' collisions, can occur),  the total cross section is given by $\sigma = 8 \pi a^2$ for identical bosons ($\sigma = 4 \pi a^2$ for non-identical particles), where $a$ is the scattering length. Elastic `s-wave' collision rates computed for the data of the experiments presented in section \ref{ExpRea} are shown in Fig. \ref{fig15} (a). During compression, the collision rate increases by almost two orders of magnitude (from $10 \, \text{s}^{-1}$ to $10^3 \, \text{s}^{-1}$). In our compressed trap, s-wave collisions are thus expected to become relevant with subsequent thermalization and evaporation.

Fig. \ref{fig15} (b) shows the cloud temperature as a function of the holding time just after compression.  The cloud is compressed during $40 \, \text{ms}$ and the trap parameters are kept constant after compression. The state of the system after compression is the initial condition for this experiment: the smaller the final trap size, the denser and hotter the cloud, leading to higher `s-wave' collision rates. On Fig. \ref{fig15} (b), we notice that the temperature reduction is faster when the cloud is initially hotter and denser (i.e. the trap size after compression is smaller), which strongly hints towards thermalization due to `s-wave' collisions. The elastic collision rates computed from Eq. (\ref{collrate}) (see caption of Fig. \ref{fig15}) are compatible with the timescale of the temperature reduction observed on Fig. \ref{fig15} (b). Indeed, three or four elastic collisions are needed for the gas to thermalize \cite{Ketterle96}. We note that pumping the atoms into the $F=1$ hyperfine level (instead of $F=2$ used in this work) would allow to avoid unwanted hyperfine changing collisions occurring at large densities.

\section{Conclusion}

We have studied a blue detuned crossed dipole trap designed to quickly compress cold atomic clouds to high densities. Extensive characterization of the system has lead to the understanding of the properties and the dynamics of this novel trapping scheme. After a very efficient loading of a large number of atoms (up to $5 \times 10^7$), the cloud is compressed in $5 \, \text{ms}$ from an initial density of $5\times10^{10} \, \text{cm}^{-3}$ to a final density of $10^{13} \, \text{cm}^{-3}$. The cloud density in the final stage is very close the Ioffe-Reggel criterion, demonstrating the efficiency and reliability of this technique to study light-matter interactions in the dense regime.

The outlook of this work includes improving the trap performances by optimizing the trap loading (e.g. denser dark MOT, compensating for gravity...), having a better control on the final compression state or using sisyphus cooling during the early compression stages. In addition to these modifications concerning the experimental protocol, improvements of the setup itself would lead to substantial performance leaps using e.g. a more detuned and powerful laser to reduce spontaneous emission losses or using VCOs with a higher input modulation bandwidth and a better frequency stability.

Increasing the trap lifetime would allow using this setup for producing quantum degenerate gases. This kind of compressible dipole trap would make the loading and evaporation proceed differently from standard approach to optical BEC, yielding larger BECs more quickly. It would consist in reaching a high collision rate during a first compression stage and then realizing runaway evaporative cooling. To do so, the trap frequency should be maintained constant by reducing the trap size when the trap power is reduced. This technique would not require mobile lenses \cite{Weiss05, BienaimePhD} allowing for faster and more stable operations.

\acknowledgments{This work was supported by ANR-06-BLAN-0096 CAROL and ANR-09-JCJC-009401 INTERLOP. We acknowledge fruitful discussions with the cold atom group at INLN. We thank Jean-Fran\c{c}ois Schaff for discussions and the imaging program.}

\end{document}